\documentclass[preprint,12pt]{elsarticle}
\usepackage{graphicx}
\usepackage{epstopdf}
\usepackage{lineno}
\usepackage{color}
\usepackage{amssymb}
\usepackage{array}
\usepackage{tabulary}
\usepackage[english]{babel}
\usepackage[T1]{fontenc}
\usepackage[latin1]{inputenc}
\usepackage{afterpage}
\usepackage{url}

\newcolumntype{C}[1]{>{\centering\arraybackslash}p{#1}}
\usepackage{booktabs}% for better rules in the table

\usepackage{multirow}
\begin{document}

\begin{frontmatter}
\title{Measurement of the cosmogenic activation of germanium detectors in EDELWEISS-III}

%% Group authors per address:
%\fnref{fn1}
\author[a]{E.~Armengaud}
\author[b]{Q.~Arnaud\fnref{fn1} }
\author[b]{C.~Augier}
\author[c]{A.~Beno\^{i}t}
\author[d]{L.~Berg\'{e}}
\author[b]{J.~Billard}
\author[e,f]{J.~Bl\"{u}mer}
\author[a]{T. de~Boissi\`{e}re}
\author[d,e]{A.~Broniatowski}
\author[c]{P.~Camus}
\author[b]{A.~Cazes}
\author[d]{M.~Chapellier}
\author[b]{F.~Charlieux}
\author[b]{M. De~J\'{e}sus}
\author[d]{L.~Dumoulin}
\author[f]{K.~Eitel}
\author[e]{N.~Foerster}
\author[b]{J.~Gascon}
\author[d]{A.~Giuliani}
\author[a]{M.~Gros}
\author[f]{L.~Hehn} 
\author[e]{G.~Heuermann}
\author[g]{Y.~Jin}
\author[b]{A.~Juillard}
\author[b,e]{C.~K\'{e}f\'{e}lian}
\author[h]{M.~Kleifges}
\author[f]{V.~Kozlov}
\author[i]{H.~Kraus}
\author[j]{V.~A.~Kudryavtsev}
\author[d]{H.~Le-Sueur}
\author[d]{S.~Marnieros}
\author[a]{X.-F.~Navick}
\author[a]{C.~Nones}
\author[d]{E.~Olivieri}
\author[k]{P.~Pari}
\author[a]{B.~Paul}
\author[d]{M.-C.~Piro\fnref{fn2} }
\author[d]{D.~Poda}
\author[b]{E.~Queguiner}
\author[l]{S.~Rozov}
\author[b]{V.~Sanglard}
\author[f]{B.~Schmidt\fnref{fn3} }
\author[e]{S.~Scorza\corref{cor1} }
\author[f]{B.~Siebenborn}
\author[h]{D.~Tcherniakhovski}
\author[b]{L.~Vagneron}
\author[h]{M.~Weber}
\author[l]{E.~Yakushev}

\address[a]{CEA Saclay, DSM/IRFU, 91191 Gif-sur-Yvette Cedex, France}
\address[b]{Univ Lyon, Universit\'{e} Claude Bernard Lyon 1,  CNRS-IN2P3, Institut de Physique Nucl\'{e}aire de Lyon, F-69622, Villeurbanne, France}
\address[c]{Institut N\'{e}el, CNRS/UJF, 25 rue des Martyrs, BP 166, 38042 Grenoble, France}
\address[d]{CSNSM, Univ. Paris-Sud, CNRS/IN2P3, Universit\'{e} Paris-Saclay, 91405 Orsay, France}
\address[e]{Karlsruher Institut f\"{u}r Technologie, Institut f\"{u}r Experimentelle Kernphysik, Gaedestr. 1, 76128 Karlsruhe, Germany}
\address[f]{Karlsruher Institut f\"{u}r Technologie, Institut f\"{u}r Kernphysik, Postfach 3640, 76021 Karlsruhe, Germany}
\address[g]{Laboratoire de Photonique et de Nanostructures, CNRS, Route de Nozay, 91460 Marcoussis, France}
\address[h]{Karlsruher Institut f\"{u}r Technologie, Institut f\"{u}r Prozessdatenverarbeitung und Elektronik, Postfach 3640, 76021 Karlsruhe, Germany}
\address[i]{University of Oxford, Department of Physics, Keble Road, Oxford OX1 3RH, UK}
\address[j]{University of Sheffield, Department of Physics and Astronomy, Sheffield, S3 7RH, UK}
\address[k]{CEA Saclay, DSM/IRAMIS, 91191 Gif-sur-Yvette Cedex, France}
\address[l]{JINR, Laboratory of Nuclear Problems, Joliot-Curie 6, 141980 Dubna, Moscow Region, Russian Federation}

\cortext[cor1]{Corresponding author: silvia.scorza@kit.edu}
\fntext[fn1]{Now at Queen's University, Kingston, Canada}
\fntext[fn2]{Now at Rensselaer Polytechnic Institute, Troy, NY, USA}
\fntext[fn3]{Now at Lawrence Berkeley National Laboratory, Berkeley, CA, USA}

\begin{abstract}

We present a measurement of the cosmogenic activation in the 
germanium cryogenic detectors of the EDELWEISS III direct dark matter search experiment.
The decay rates measured in detectors with different exposures to cosmic rays
above ground are converted into production rates of
different isotopes.
The measured production rates in units of nuclei/kg/day are 
82 $\pm$ 21 for $^3$H,
2.8 $\pm$ 0.6  for $^{49}$V,
4.6 $\pm$ 0.7 for $^{55}$Fe, and
106 $\pm$ 13 for $^{65}$Zn. 
These results are the most accurate for these isotopes. 
A lower limit on the production rate of $^{68}$Ge of 74~nuclei/kg/day is also presented. 
They are compared to model predictions present in literature and 
to estimates calculated with the ACTIVIA code.

\end{abstract}
\begin{keyword}
%% keywords here, in the form: keyword \sep keyword
Germanium detector \sep  Cosmogenic activation \sep Dark matter detection %\sep Double-beta decay
%% PACS codes here, in the form: \PACS code \sep code

%% MSC codes here, in the form: \MSC code \sep code
%% or \MSC[2008] code \sep code (2000 is the default)

\end{keyword}

\end{frontmatter}
%%%%%%%%%%%%%%%%%%%%%%%%%%%%%%%%%%%%%%%%
%\tableofcontents
%%%%%%%%%%%%%%%%%%%%%%%%%%%%%%%%%%%%%%%%
\section{Introduction}

Germanium is widely used as a detector material in experiments searching for a rare process
like the interaction of weakly interacting massive particles (WIMPs)~\cite{dmreview}.
It is possible to build detectors with very good energy resolution
based on the measurement of the ionization produced in the particle interaction,
or of the increase of temperature~\cite{gebolo}.
In addition, the combination of the ionization and heat signals is a powerful tool to distinguish
nuclear recoils from electron recoils.
Moreover, the crystal-growing process used in the semiconductor industry 
purifies the material to a high level that matches well the stringent ultra-low radioactivity
requirements of rare event searches.

The potential of germanium detectors for achieving very low threshold below 1~keV
 is particularly
attractive for searches of WIMPs with masses below 10~GeV/c$^{2}$.
The background at energies below 20~keV in such a detector is thus of particular interest.
Notably, the contribution from tritium beta decays may have a significant impact
on the sensitivity of the next generation of these detectors.

The crystallization process removes all cosmogenically-produced radioactive atoms, 
with the exception of unstable germanium isotopes like $^{68}$Ge (see below).
Their populations grow back again when the crystal is kept above ground, 
and therefore exposed to cosmic rays and the associated hadronic showers.
Short-lived isotopes decay rapidly as soon as the detectors are stored underground, 
deep enough to suppress the hadronic component of the cosmic rays~\cite{Farley2006451}.
The isotopes that merit attention have lifetimes exceeding a few tens of days,
since shorter-lived nuclei can be eliminated just by storing the detectors in the underground
site for some reasonable time before starting data taking.

The cosmogenic products that have the most noticeable effect on the low-energy
spectrum recorded in germanium detectors are those that decay via electronic capture (EC). 
The capture is often followed by the emission of a $K$-shell X-ray with characteristic energy between
4 and 11~keV. 
$L$- and $M$-shell captures will produce weaker lines at approximately 1 and 0.1~keV, respectively.
The sharp line shapes and known $K$:$L$:$M$ intensity ratios can be used to 
identify and subtract the associated events.
However, it is preferable to reduce their initial intensities to the lowest possible level.
Measurements of the production rates of EC-decaying isotopes
is helpful in designing a detector-production procedure that limits these
backgrounds to acceptable levels, and, more generally, to constrain models
predicting the production rates of all isotopes, including those that may prove
to elude direct measurements.

Another type of background of particular interest is the beta decay of tritium ($^{3}$H) originated from nuclear 
reactions induced by the interaction of the hadronic component of cosmic rays with atoms in the material~\cite{avignone}.
The electron emitted in the beta decay
of tritium has an end point $Q_{\beta}$ of only 18.6~keV, and thus
contributes to the background of low-energy events over the entire energy range
relevant for low-mass WIMP searches.
The lifetime of $^{3}$H is particularly long ($\tau$ = 17.79~y), 
so the tritium activity can essentially be expected to remain almost the same 
throughout the life of the detector.
The only way to reduce this background is to limit the exposure of the crystal
between the time it is grown and its installation underground.
There are large uncertainties in model predictions for 
the production rate of $^{3}$H, 
and available measurements can only provide crude upper limits~\cite{mei}.

The EDELWEISS collaboration has operated an array of 24 
germanium heat-and-ionization detectors with the objective to perform 
searches for WIMPs with a total exposure close to 3000~kg$\cdot$d,
and more specific searches for low-mass WIMPs with a subset of
its detectors with the best experimental energy thresholds~\cite{lowmass}.
The experiment is located in the Laboratoire Souterrain de Modane
(LSM) and protected by a mean rock overburden of 1800~m (4800~m.w.e.)
that reduces the cosmic ray flux to about 5~$\mu$/m$^2$/day~\cite{Schmidt:2013gdc}, 
i.e.\ 10$^6$ times less than at the surface.
The detectors are covered by interleaved electrodes
that provide an efficient tool to reject surface events
(i.e.\ particle interactions taking place within $\sim$2 mm from the detector surface) 
down to energies of $\sim$1~keV~\cite{lowmass}.
The resolutions achieved with these detectors, 
the reduction of the external gamma-ray background 
and the excellent surface-event rejection performance of the interleaved 
electrodes~\cite{broniatowski}, 
have made possible a precise measurement
of decay rates of different nuclei in the bulk volume of
germanium detectors, and in particular, for the first time, to
measure unambiguously the intensity of the tritium spectrum.
Efforts were made to keep to a minimum the exposure of
each crystal to cosmic rays throughout the detector production.
A history of the key steps in the detector production process is available.
Despite this, there are non-negligible systematic uncertainties in the recorded history of exposure times.
However, these uncertainties can be tested, because unforeseen production delays\footnote{
These delays occurred to solve problems related to surface current leakage, as described in
\protect{\cite{leakage}}.} 
led to a relatively large  spread (up to a factor of 4) in the exposures of the
different detectors to cosmic rays.
It was therefore possible to check on isotopes with the largest statistics 
that the observed activation rates scaled with
the expectations from the recorded history of exposure times.

In the following,
we detail the EDELWEISS-III setup relevant
to this measurement (Section~\ref{sec:setup}), 
as well as the data selection (Section~\ref{sec:data}).
We present the expected properties of the activation of
tritium and other isotopes and of the energy spectrum of the emitted electrons,
and describe the analysis used to extract their intensities from the data~(Section \ref{sec:analysis}).
These results are then converted into production rates during the exposure above ground
(Section~\ref{sec-history}) and compared to a previous measurement and calculations
(Section~\ref{sec:pr}).

\section{Experimental setup}\label{sec:setup}

The active target of the EDELWEISS-III WIMP search experiment consists of 
twenty-four 800-g {\em Fully-InterDigit} (FID) germanium detectors, cooled down to an 
operating temperature of 18 mK.
All materials surrounding the detectors have been tested for their radiopurity~\cite{scorza}.
The cryostat is surrounded by 20~cm of lead and 50~cm of polyethylene shielding.
A constant flow of de-radonized air in the vicinity of the cryostat reduces the radon 
level down to 30 mBq/m$^3$. 
The shielding is surrounded by a 100 m$^2$  muon veto made of
plastic scintillator modules
with a geometrical coverage of more than 98\%~\cite{Schmidt:2013gdc}.

The detectors are high-purity germanium cylindrical crystals of 70 mm in diameter
and 40 mm in height. 
All surfaces are entirely covered with interleaved Al electrodes, 
biased at alternate values of potentials. 
The electrodes are 150 $\mu$m in width and separated by 2 mm.
The potentials applied to the electrodes are chosen  to determine
an axial electric field in the bulk of the detector~\cite{broniatowski}, 
while in the volume within about 2~mm from the surfaces
the electric field is parallel to them. 
As a consequence, electron-hole pairs created in the bulk volume are 
collected in the axial field by the fiducial electrodes on both sides of the detector, 
while surface events will be collected by adjacent electrodes.
This scheme extends also to the cylindrical surfaces.
Fiducial (or bulk) events  can thus be selected on the basis of the presence
of signals of opposite signs on the fiducial electrodes on each side of the detector,
and on the absence of signals on all other electrodes.
The fiducial ionization $E_{fid}$ is defined as the average of the signals on the two fiducial electrodes.
Calibrations with a $^{210}$Pb source in equilibrium with its Bi and Po daughters
have shown that this technique can reject surface $\beta$'s, $\alpha$'s and 
$^{206}$Pb recoils with an efficiency of the order of 99.999\%~\cite{surfacerejection}.

Nuclear recoils are identified by comparing the ionization signal
with the $\sim\mu K$ rise in temperature accompanying each interaction,
measured with two 15~mm$^{3}$ germanium NTD (Neutron Transmutation Doped) thermistors glued on each
side of the detector.
The charge and temperature signals are calibrated using a $^{133}$Ba
$\gamma$-ray source: 
the units of both signals are thus in keV-electron-equivalent (keV$_{ee}$).
Since this work is devoted to the study of bulk electron recoil populations,
the subscript $ee$ will be omitted in the following.
The signals on both thermistors are combined into a single heat measurement $E_c$.

The average dispersion in ionization signals $E_{fid}$ at 0 keV  
is characterized by $\sigma_{fid}$ = 230 eV (baseline resolution).
For the heat signal $E_c$, the corresponding values of $\sigma_{c}$
range from 150 to 500~eV depending on detectors.
The resolution increases with the energy of the signal, with a linear term
that is dominated by charge trapping effects~\cite{quentin}. 
At 356 keV, the resolutions $\sigma(E)$ are larger than the 
baseline $\sigma_{fid}$ and $\sigma_{c}$ values
and are approximately 4 keV.

The data acquisition triggers if the amplitude of one of the heat signals 
is above a threshold value\footnote{The online 
filtering on the heat signals is less efficient than the offline version,
resulting in a slight degradation of resolution. Here, $E_c$ and $\sigma_{c}$
refer to the offline-filtered heat signals. 
As the trigger is applied to the signals calculated online, 
some $E_c$ values may be less than their corresponding
online trigger threshold by as much as $\sigma_{c}$.}. 
These values are automatically adjusted every minute for each heat sensor 
according to the event rates recorded minute by minute.
This, together with the 0.5 kHz sampling of the heat channels, results
in the possibility to set the thresholds to levels that are close
to $\sim$4.5$\sigma_{c}$, while keeping
the trigger rates per detector at approximately 50 mHz.

\section{Data selection}\label{sec:data}

\subsection{Data set and detector selection}

The data was recorded over a period of 280 calendar days,
from July 2014 to April 2015.
During that period, 160 days were devoted to WIMP searches.
All twenty-four detectors were used to define coincidences between
detectors.
The average heat trigger threshold and baseline resolutions were monitored hour by hour.
For this analysis, only hours when this trigger threshold is below 2 keV were selected.
The hourly fiducial baseline resolution on the fiducial ionization measurement 
was also required to be less than 400 eV.
Two detectors with a failing ionization channel, preventing an efficient application
of the fiducial cut, were excluded from this exposure.
Three detectors had less than one day of running time with an online 
threshold of below 2 keV and were also excluded from further analysis.
The dead-time corrected exposure of the remaining 19 detectors is 
1853 detector$\cdot$day.
This exposure is considered for different global fits used in order to evaluate systematic 
uncertainties.
As each detector had a different history of exposure to cosmic rays,
the final results will be based on the fits to individual detector data for the 13 detectors
with individual exposure greater than 60 days, 
corresponding to 87.0\% of the total exposure.

Similarly, a sub-sample of events with an online threshold below 0.8 keV is also
defined, for precision tests of the efficiency correction and of the sample
purity at energies lower than used in the final analysis. This sample
corresponds to 499 detector$\cdot$day, to which 10 detectors contribute.

\subsection{Event selection}

\label{sect-neg}
An event is included in the analysis if its  $E_{fid}$ ($E_c$) value
is larger than 3.5$\sigma_{fid}$ ($3\sigma_{c}$).
The values of $\sigma_{fid}$ and  $\sigma_{c}$ are measured hour by hour
from Gaussian fits to the distribution of amplitudes observed in
events that have not triggered the online threshold.
Cuts are also performed on the $\chi^2$ of the fit of template pulse shapes to 
the ionization signals, resulting in an efficiency loss of less than 1\%.

\begin{figure}
\begin{center} \includegraphics[scale=0.72]{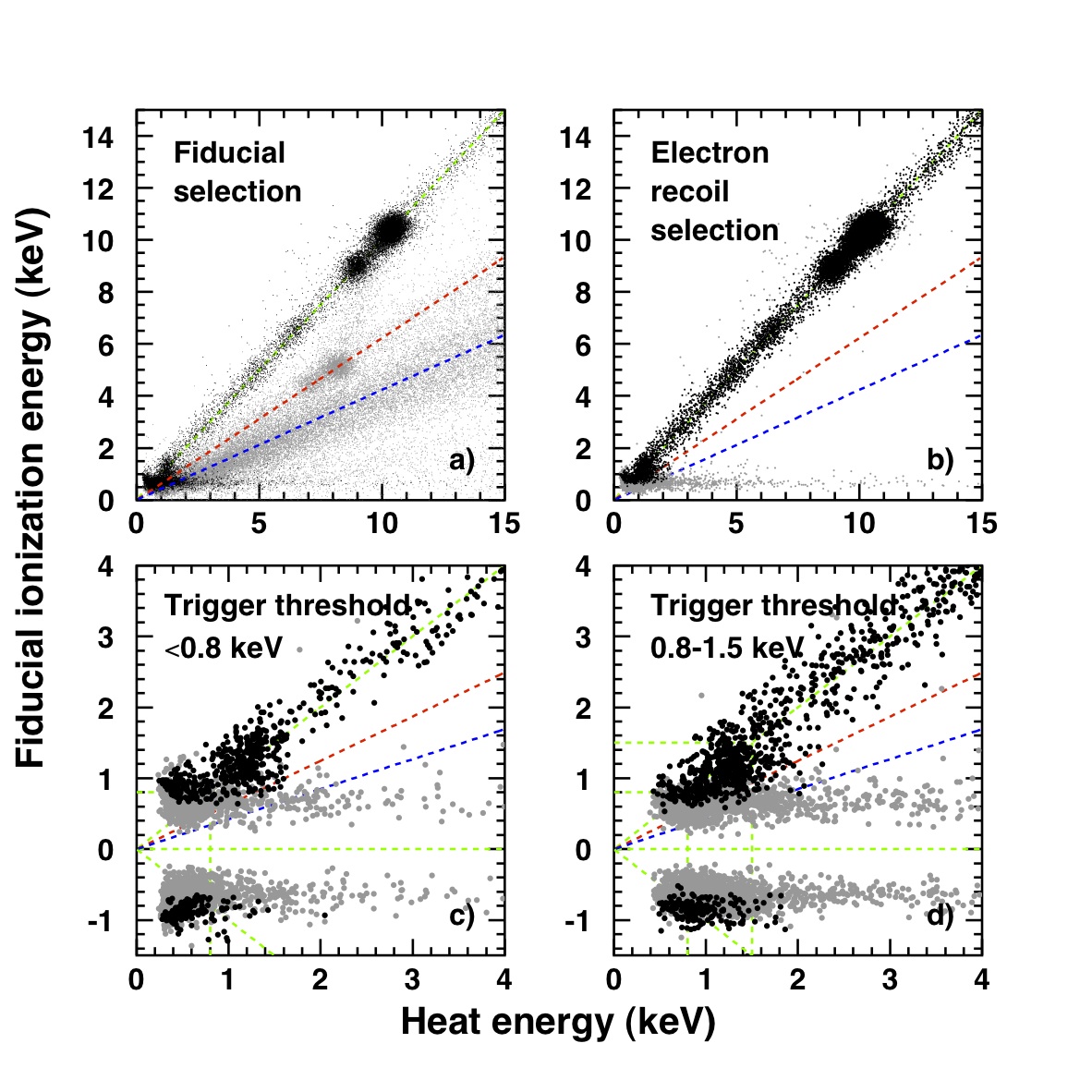} \end{center}
\caption{
Distribution of the fiducial ionization $E_{fid}$ versus heat energy $E_c$ at different steps of the data selection. 
Panel a): events passing the $\chi^{2}$ selection, and having $E_c>3\sigma_{c}$ and $|E_{fid}|>3\sigma_{fid}$,
where the black markers are events passing the fiducial selection, and those in gray do not pass this cut.
Dotted red and blue lines represent the average ionization yields of surface gamma and surface
beta events, respectively. 
Panel b): events after the fiducial selection, with black markers representing the events passing the $E_c-E_{fid}$ consistency cut
to select fiducial electron recoils, and gray markers to represent those that do not pass this cut.
Bottom panels: same as figure b), but now extended to negative ionization energies
to better illustrate the contamination by heat-only events (see text).
Figures c)  and d) refer to events with a
trigger selection on the heat signal (as measured online) below 0.8 keV, 
and in the range 0.8-1.5~keV, respectively. 
\label{fig-datasel}}
\end{figure}

Fig.~\ref{fig-datasel}a) shows the distribution of the fiducial ionization $E_{fid}$
as a function of the heat signal $E_c$ for selected events in a sample where the cut on
$E_{fid}$ has been relaxed from $3.5\sigma_{fid}$ to $3\sigma_{fid}$.
On this figure, events in gray are those rejected by the fiducial cut.
Namely, events are considered as having occurred in the fiducial volume if the
signals on the two non-fiducial electrodes, as well as the difference between
the two fiducial electrodes, are each consistent with noise within $\pm$2.574$\sigma$ (99\%).
The rejected events  are mostly distributed along the two dashed lines 
on Fig.~\ref{fig-datasel}a) representing
the expected location for surface beta events (blue dotted line) and
surface gamma events (red dotted line).
Fig.~\ref{fig-datasel}b) shows the same distribution after the above-described
selection of fiducial events.
The events in gray are those where the values of $E_{fid}$ and $E_c$ are
not consistent with each other within 2.574$\sigma$ of the experimental
resolutions.
They are associated with the background of heat-only events discussed
in Ref.~\cite{lowmass}.
Their origin is illustrated in Figs.~\ref{fig-datasel}c) and d), where the low-energy
part is shown together with events selected with cuts calculated using $|E_{fid}|$ 
instead of $E_{fid}$, 
namely, $|E_{fid}|$ $>$ 3$\sigma_{fid}$ and the consistency cut is applied on ($E_c-|E_{fid}|$).
For clarity, events with trigger threshold below 0.8 keV (Fig.~\ref{fig-datasel}c) are
shown separately from those with trigger thresholds between 0.8 and 1.5~keV
(Fig.~\ref{fig-datasel}d).
The distribution of $E_{fid}$ for heat-only events is a Gaussian centered at $E_{fid}=0$.
In Figs.~\ref{fig-datasel}c) and d), 
what is therefore observed after the rejection of events with $|E_{fid}| <3 \sigma_{fid}$
are two populations of events, symmetric in $\pm~E_{fid}$.
A leakage from the population of surface gamma and beta events would be centered
around the red and blue dotted lines, respectively, on Fig.~\ref{fig-datasel}c) and d),
and not distributed symmetrically at $\pm E_{fid}$. 
Such a symmetry is clearly observed for the events in gray, corresponding to those
rejected by the consistent cut on $E_{c}-|E_{fid}|$.
Most of the potential backgrounds below 2 keV thus comes from heat-only events.
The population of heat-only events with $E_{fid}>0$ that are not rejected
by the consistency cut will be estimated by mirroring the equivalent
population with $E_{fid}<0$.
 For instance, in Fig.~\ref{fig-datasel}c) and for $E_c<$0.8 keV, 
 the population of gray points
 with $E_{fid}<0$ is nearly equal to those
 with $E_{fid}>0$. 
 This is also true for the events in black, indicating that most events 
 with positive ionization energies and $E_c$ values below 0.8 keV
 remaining after applying the consistency cut 
 are due to the tail of heat-only events.
Figure~\ref{fig-datasel}c) suggests that subtracting a sideband
corresponding to events with $E_{fid}<$0
should confirm the clear observation of the peak of events at $\sim$1.2~keV due to
the capture of $L$-shell electrons in Ga, Ge and Zn atoms (Table~\ref{tab-decay}).
The efficiency of the entire analysis procedure at low energies has been tested 
(see Section~\ref{sec-yield})
by comparing the $L/K$ intensity ratio obtained with the known theoretical values,
in the data sample with an online trigger threshold below 0.8 keV.
It can also be observed on Fig.~\ref{fig-datasel}c) that the 1.2 keV peak
may extend to the red dashed line corresponding to surface gamma events.
This effect  will be taken into account when calculating
the energy dependence of the efficiency in Section~\ref{sec-eff}.
It can also be observed from Fig.~\ref{fig-datasel} that the background
above 2 keV is expected to be very small.

\subsection{Selection of multiple-hit events}

\label{sec-coinc}
As can be deduced from the electron capture decay data listed in Table~\ref{tab-decay},
events associated with $^{49}$V, $^{55}$Fe, $^{68}$Ga and $^{68}$Ge
are expected to produce an $L$- or $K$-shell energy deposit in a single FID detector.
The 18.6 keV electron emitted in the beta decay of tritium is also expected 
not to escape the fiducial volume of the detector where it occurs.
Backgrounds unrelated to these decays can be reduced by rejecting events where
more than one detector have triggered simultaneously.
An event is considered as a single if no other detectors have triggered within an
interval of 10~ms, or if the sum of the ionization energies of all other detectors having
triggered in this interval is less than 1~keV.

However, Table~\ref{tab-decay} also shows that
events associated with $^{65}$Zn (and $^{54}$Mn) are accompanied
by the emission of a $\gamma$ ray with an energy of 1115.5~keV  (834.9~keV)
with a branching ratio of 50.0\%  (100\%).
This $\gamma$ ray (or internal conversion electron) may be detected simultaneously 
in the same detector, in which case no $L$- or $K$-shell peaks are observed.
A high-energy $\gamma$ ray can also escape the FID fiducial volume 
and be detected in one of the 23 other detectors.
Thus, some of the $^{54}$Mn and $^{65}$Zn decays are associated with
coincident events.
As a result, the line intensity for these events will be obtained by a simultaneous
fit of the spectra of single- and multiple-hit events.
In addition, the intensity for $^{65}$Zn  will be corrected 
for the fraction of unobserved events at 8.98 keV due to the absorption of the
1115.5~keV $\gamma$-ray inside the detector itself, as estimated
by Monte Carlo simulation, accounting for  25$\pm$1\% of all $^{65}$Zn decays. 
The $^{54}$Mn line is included in the peak model, 
but its intensity is too weak to extract a reliable rate. 

\subsection{Energy spectrum}

\label{sec-eff}
The energy of an event is taken as the average of $E_c$ and $E_{fid}$
weighted by the inverse of the square of their associated baseline resolutions 
$\sigma_{c}$ and $\sigma_{fid}$, as measured hour-by-hour.
A side-band spectrum corresponding to the contribution of heat-only events,
obtained by replacing $E_{fid}$ by $-E_{fid}$ in the weighted average,
as described in Section~\ref{sect-neg}, has been subtracted from the resulting spectrum. 
The spectrum is then corrected for the energy dependence of
the efficiency due to 
{\em i)} the fiducial volume cut,
{\em ii)} the cut on $E_{fid}>$3.5$\sigma_{fid}$, and
{\em iii)} the online trigger threshold on the heat energy\footnote{
The cut on $E_c>$3$\sigma_{c}$ has a negligible effect on the efficiency
once the effect of the online trigger threshold is taken into account.}.
The fiducial mass value per detector~\cite{lowmass} is measured with the $K$-line
intensities, and thus effectively includes an
inefficiency due to the rejection of events where the
apparent signal on a given electrode comes from a gaussian
fluctuation of its baseline noise.
As the energy of the event approaches zero keV,
the expected signal on the non-fiducial electrodes for
surface events becomes smaller than the baseline noise,
and the fiducial cut  can no longer reject them.
The selected volume becomes the entire detector,
and the effective volume increases from 70\% to 100\% of the detector.
Fig.~\ref{fig-datasel}c) and d) hint at the presence of this effect as
the distribution of events around the peak at 1.2 keV extends out to the line where 
surface gamma rays are expected.
This energy dependence is calculated using the measured baseline 
resolutions on the signal on the non-fiducial electrodes and 
on the difference between those on the two fiducial electrodes.
This efficiency correction is applied to the detector energy spectra, 
but as a convention, 
the fiducial exposures in all the following will be quoted for
a $\gamma$ ray of 10 keV.
The energy-dependence of the efficiency induced by 
the 3.5$\sigma_{fid}$ cut on $E_{fid}$ and
the online heat threshold
have been calculated for each detector using the hour-by-hour
measurements of both values, and taking into account the smearing
effect due to the difference between the online and offline
reconstruction of $E_c$.

The calculated corrections are very close to 1 over the entire energy range
down to the analysis threshold, set at 2 keV.
The validity of these corrections are thoroughly tested by applying them
to the sample of events where the online threshold is below 0.8 keV.
The test consists of a comparison of the measured and theoretical $L/K$ intensity ratio 
for this data sample.
The model predicts that the $L/K$ efficiency ratio varies from 0.92 to 1.01 between 1.1 and 1.3 keV (the energies of the $L$-shell
lines for $^{65}$Zn and $^{68}$Ge, respectively: see Table~\ref{tab-decay}).

The resulting efficiency-corrected spectrum for the 1853 detector$\cdot$day sample 
is shown in Fig.~\ref{fig-spectrum}.
The inset shows the efficiency-corrected spectrum for
the data set with an online trigger threshold below 0.8 keV (499 detector$\cdot$day), 
used to test the efficiency model and the heat-only sideband correction
down to 1 keV. 
The electron capture and the tritium beta-decay intensities described below 
are taken from the fit of the sample of events with the 2 keV threshold cut.

\section{Decay rate measurements}\label{sec:analysis}

The decay rates of the different cosmogenically produced isotopes 
are obtained from simultaneous fits  of the energy spectra of
single and multiple hit events, as defined in Section~\ref{sec-coinc}.
The fit is first applied to the co-added spectra of Fig.~\ref{fig-spectrum} in order
to test its validity and estimate global systematics, and then to individual
detector spectra in order to obtain individual count rates per kilogram and per day 
for each of the isotopes of interest.
The fit model has three components: 
{\em i)} the tritium spectrum, 
{\em ii)} a Compton background and
{\em iii)} $K$- and $L$- spectral lines.

\subsection{Tritium beta spectrum}

The energy spectrum of emitted electrons has an end-point value of $Q_{\beta}$~=~18.6 keV. It is described by~\cite{Brown:1995xn}: 

\begin{equation}
\frac{dN}{dt} \propto \sqrt{T^2+2mc^2T} \; (T+mc^2) \; (Q_{\beta}-T)^2 \; F(T,Z=2)
\end{equation}

\noindent where $T$ is the kinetic energy, $m$ is the mass of the electron, 
$c$ the speed of light 
and \mbox{$F(T,Z=2)$} is the Fermi function
for tritium decay.
This function can be approximated in the non-relativistic limit as 
$x/(1-e^{-x})$, where $x=\frac{4\pi\alpha c}{v}$, $\alpha$ is the fine structure
constant and $v$ is the electron velocity.
With this Fermi function, and $T$ expressed in keV, the spectral shape becomes:

\begin{equation}
\frac{dN}{dt} \propto (T+mc^2) \; (Q_{\beta}-T)^2 \;
                                (1-e^{-\frac{1.466}{\sqrt{T}}})^{-1}
                                \label{eqn-shape}
\end{equation}

This function is used to describe the tritium component in the fit.
The tritium intensity is taken as the integral of the fitted component from zero to 18.6 keV.
The spectral shape of Eq.~(2) was not smeared with the detector-dependent 
energy resolutions, as it was verified that this procedure has an impact only
below the analysis threshold, and in a region around the end point that has a negligible statistical 
weight in the determination of the intensity.

\subsection{Compton background}

\begin{figure}
\begin{center} \includegraphics[scale=0.75]{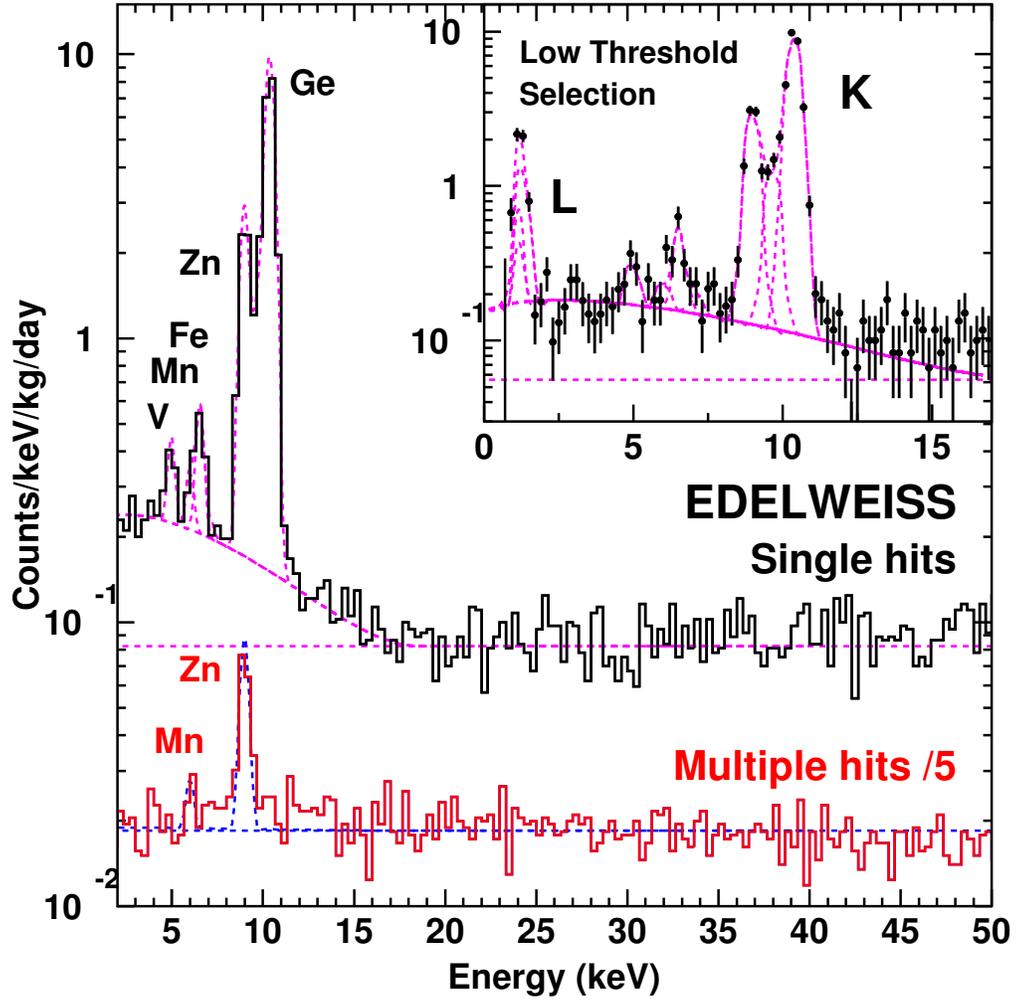} \end{center}
\caption{Efficiency-corrected spectrum for the 1853 detector$\cdot$day sample,
together with the fit to the data described in the text.
The distribution for single events is in black, and the red spectrum representing
multiple-hit events has been scaled down by a factor of 5 for clarity.
The inset shows the efficiency-corrected spectrum for
the 0.8 keV selection (499 detector$\cdot$day), used to test down to 1 keV the
efficiency model and sideband correction described in the text and in Fig.~\protect{\ref{fig-datasel}}.\label{fig-spectrum}}
\end{figure}

All the peaks visible below 12~keV  on Fig.~\ref{fig-spectrum} 
are due to the electron capture decays of
different isotopes (mainly Ge, Ga and Zn, see Table~\ref{tab-decay}) 
that will be detailed in Section~4.3.
Between 20 and 50~keV, the Compton background is constant within statistical errors,
with an average of 0.090 $\pm$ 0.002~counts/kg/day/keV.
Studies of the Compton plateau below 50 keV in $^{133}$Ba
calibrations and GEANT4 Monte Carlo simulations also indicate that this
flat behavior between 20 and 50 keV does not hint
%not exclude the existence
to a possible slope below 20 keV.  
Nevertheless, variations as much as $\sim$10\% of the Compton rate between 0 and  20 keV cannot 
be excluded, but a similar slope should also be present in multiple-hit events.
The energy spectrum of multiple-hit events is also shown on Fig.~\ref{fig-spectrum}. 
Apart from the expected contributions of  the $^{65}$Zn and, 
possibly, $^{54}$Mn peaks (discussed below), 
the spectrum between 2 and 50 keV appears to be relatively
consistent with a flat Compton plateau.
The statistical significance of this sample is sufficient to provide an
alternative template for the shape of this background.
It can be noted that rejecting multiple events reduces the Compton background by
almost a factor of two while having no effect on tritium decays.

In the following, the tritium and EC line intensities are obtained from a fit 
that includes a flat Compton background.
As the fitted tritium intensity is somewhat more sensitive to the assumed shape
of the Compton continuum, 
a systematic uncertainty is taken into account as the difference between this result
and the intensity obtained with different background assumptions,
namely {\em i)} a first order polynomial and {\em ii)} 
the multiple-hit spectrum scaled by the ratio of the number of counts between 20 and 40~keV in the single and multiple-hit spectra.
The first test is performed for each individual detector. 
The second test requires more statistics and is only applied to the spectrum
of Fig.~\ref{fig-spectrum}.

\subsection{Electron capture lines}

The model for the peaks in the spectrum
is not only important for getting the EC decay intensities but also
to constrain the amplitude of the underlying tritium component.
The structure of peaks considered in the fit is the following.
The energy-dependent width of a peak is given by 
$\sigma(E)$ = $\sqrt{\sigma^2_0+(\alpha E)^2}$,
where  $\alpha$ is a free parameter and $\sigma_0$ is fixed to the value derived from
the measured average baseline resolutions $\sigma_{fid}$  
and $\sigma_{c}$:
$\sigma_0=1/\sqrt{\sigma^{-2}_{fid}+\sigma^{-2}_{c}}$.
The global energy scale is also a free parameter of the fit.

Table~\ref{tab-decay} lists all the electron capture (EC) decays with $t_{1/2}>$ 10~days,
and their daughter decays, for 21$\leq Z \leq$33. All these can potentially lead to
peaks at the $K$-shell binding energy $E_K$ in a germanium detector.
The detectors are fully efficient to electrons, positrons and all forms of radiation down
to a few eV. 
The energies of all prompt ($<1$~ms) cascades of radiation are summed up
into the same event, 
so only pure EC decays can produce a peak at energy $E_K$.
By applying the fiducial selection, we ensure that the radiation is emitted 
at least a few
millimeters below the detector surface, and only $\gamma$ rays with energies
above some hundred keV can escape from being summed up with the X-ray energies.

Since the detectors have been installed at LSM between 6 and 18 months before
the run start in July 2014, only isotopes with half-lives longer than 100 days need
to be considered.
An exception is $^{71}$Ge. This nucleus has a half-life of 11~days,
but it is produced by $(n,\gamma)$ absorption
when the detector is exposed to a neutron source. 
This decay produces a peak at the same energy $E_K$ = 10.37~keV
as the EC decay of $^{68}$Ge.
Consequently, 
some care must be taken to exclude data following neutron calibrations
when extracting the $^{68}$Ge line intensity. 

The EC decay of $^{68}$Ge produces a daughter with a half-life
$t_{1/2}$ = 67.7~min,  
$^{68}$Ga, that can also undergo an EC decay to $^{68}$Zn with
a line at $E_K$ = 9.66~keV 
with a branching ratio of 11.12\%.
This value should correspond to the  ratio of the peak intensities 
from $^{68}$Ga and $^{68}$Ge EC decays. 

Other observable peaks should be those at the $E_K$ values of the EC decay of
$^{65}$Zn, 
$^{55}$Fe, 
$^{54}$Mn and
$^{49}$V.
Therefore, peaks at those energies are included in the fit, together
with those at 9.66 and 10.37~keV.
No peak associated to the EC decay of $^{57}$Co is expected
to be observed  at $E_K$ = 7.11~keV in our detectors, 
since this decay is accompanied with 
122 and 136~keV $\gamma$-rays that are easily absorbed 
inside the fiducial volume and the $\sim$2~mm layer of Ge that surrounds it.
The same is true for $^{44}$Ti decays, unobservable in the fiducial
volume of the detector because
of the accompanying 78.3 and 67.9~keV transitions.
However, the EC decay of its daughter $^{44}$Sc, with a half-life
$t_{1/2}$ = 4.0~h,
is in principle observable at $E_K$ = 4.04~keV, 
as the accompanying 1157.0~keV $\gamma$-ray
may escape detection inside the detector.
The total spectrum of 
Fig.~\ref{fig-spectrum} was fitted including peaks
at $E_K$ = 4.04 and 7.11 keV.
The resulting intensities were consistent with zero within their statistical uncertainties,
and had no significant impact on the fitted tritium intensities.
The heights of these two peaks have been fixed to zero for all
subsequent fits.

The $L$-shell peak structure only needs to be included in the fit of the data sample
with the $<$0.8~keV online threshold requirement (inset of Fig.~\ref{fig-spectrum}). 
This multiplet structure is included in the fit as peaks at 1.10, 1.19 and 1.30~keV
(Table~\ref{tab-decay}) 
with intensities taken as those of the peaks at 8.98, 9.66 and 10.37~keV
scaled by the same $L/K$ ratio, which is a free parameter of the fit.

\subsection{Fit to the spectra and uncertainties}

The simultaneous fit to the single- and multiple-hit data shown
in Fig.~\ref{fig-spectrum} have relatively good $\chi^2/d.o.f.$ values 
(1.12 and 1.03 for the fit to the data with an online trigger threshold below
2 and 0.8~keV, respectively).
The fitted number of tritium counts varies by 1\% whether the Compton background
is assumed to be flat or to be a first order polynomial.
Changing the binning of the data has a similar effect.
A larger difference appears if the Compton shape is taken from that of
the multiple-hit spectra instead of being assumed to be flat.
In this case the fitted number of tritium counts decreases by 11\%. 
This value is considered as a systematic uncertainty of the 
fitted tritium intensity associated with the uncertainty in the shape of the
Compton background spectrum at low energy.

The fitted $L/K$ ratio of the intensities of the peaks 
shown in the inset of Fig.~\ref{fig-spectrum} is
0.113 $\pm$ 0.008 (statistical error only), 
and agrees well with the theoretical value of 0.11~\cite{PhysRev.132.362}.
It can also be interpreted that the efficiency correction recovers at least
93\% (90\% C.L.) of the full efficiency at an energy of 1.25 keV, 
combination of the three $E_L$ peaks of Table~\ref{tab-decay}.
The efficiency model described in Section~\ref{sec-eff}
is more robust for energies above 2.0 keV,
as turning on or off entirely each of the three elements in the efficiency model
changes the fitted tritium intensity by less than 2\%.
The associated systematics is neglected as it is much smaller than
the 11\% uncertainty described in the previous paragraph.

\subsection{Activity measurements}
\label{sec-yield}

The fit of the entire 2~keV threshold data set results in an observed tritium activity of 
0.94~$\pm$ ~0.06~(stat.)~$\pm$ ~0.10 (syst.)~events
per day and per detector, extrapolated down to 0~keV using Eq.~(\ref{eqn-shape}).
However, the rate can be expected to vary from detector to detector.
The fit is therefore performed for each detector individually.
The tritium rate is obtained from the fit with the flat Compton background.
The systematic uncertainty associated with the Compton shape is
taken as the quadratic sum of the common 11\% contribution discussed in the 
previous section
and the difference between the fitted value with the flat and linear Compton backgrounds
in individual detectors.
The same systematic tests applied to the peak intensities
showed that they are not significantly affected by uncertainties in the
shape of the Compton background.

\begin{figure}
\begin{center} \includegraphics[scale=0.75]{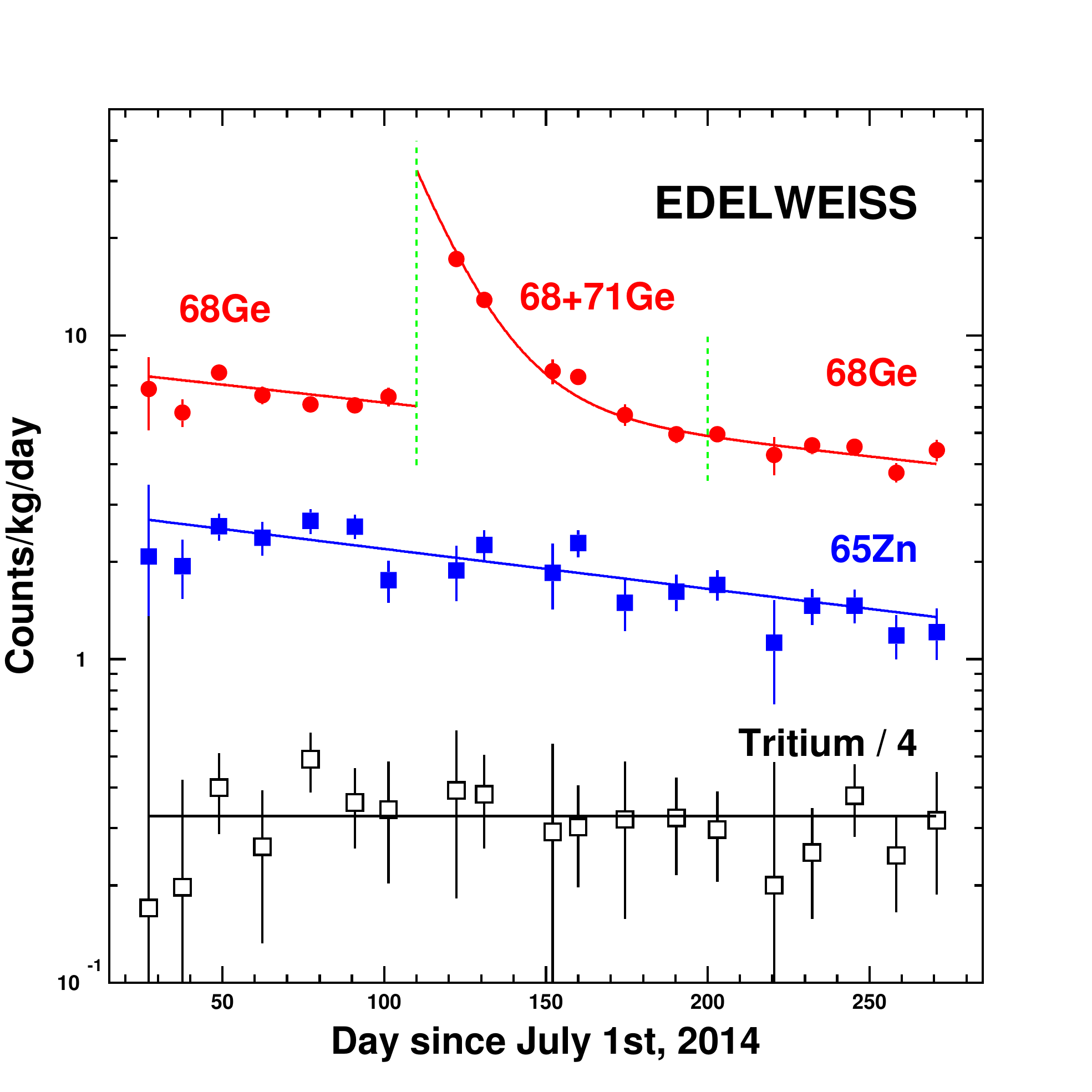} \end{center}
\caption{
Count rates (in counts per kg$\cdot$d) in the peaks at 8.98~keV (Zn) and 10.37~keV (Ge) 
as well as in the tritium spectrum, 
as a function of the time (in days since July 1$^{st}$, 2014). Each rate and the average 
time at which it was measured is calculated in bins of 14 days,
and excludes the data set taken with calibration sources.
The tritium rates have been scaled by a factor 4 for clarity.
The full lines are fits to the data of exponentials with the expected
lifetime for each isotope. The vertical dashed lines represent
the data set excluded from the $^{68}$Ge rate measurement,
as it is clearly affected by the neutron activation of germanium
following the AmBe calibration occurring between days 108 and 112.
\label{fig-time}}
\end{figure}

The $^{68}$Ge rate is obtained from a fit that excludes the 90 days
following the calibration of the detector with an AmBe neutron source.
Fig.~\ref{fig-time} shows that this cut should essentially 
remove all contaminations of the $^{68}$Ge peak by $^{71}$Ge decays.
The $^{68}$Ga/$^{68}$Ge ratio derived from the data outside this
interval is  0.117 $\pm$ 0.006 and is compatible with the
value of 0.1112 from Table~\ref{tab-decay}.

Fig.~\ref{fig-time} also shows that the decay rates of the $^{68,71}$Ge
and $^{65}$Zn decrease with the proper time constant,
while the tritium rate is constant, as expected from its very long half-life.
As the total length of the data taking period is close to one year,
the decay rate of isotopes with half-lives $t_{1/2}$ <~3~y can not be considered constant.
%As the total length of the data taking period is close to one year,
%the measured average activity of decays with half-lives $t_{1/2}$ <~3~y
%corresponds to the rate 
%at a time which is related to the half-life itself, 
%and therefore 
%depends on the nucleus under consideration.
%To avoid this arbitrariness, 
Then, the activity of an isotope with lifetime
$\tau= t_{1/2} / \ln 2$ has been obtained from a fit where each
event is weighted by $\exp(\frac{t-t_{ref}}{\tau})$,
where $t_{ref}$ corresponds to the middle of the data-taking period, 
i.e.\ October 30$^{th}$, 2014.
The fitted activity is thus the activity on this date.

The resulting decay rates per detector are listed in Table~\ref{tab:rates}.
The measured rates are expressed in terms of events per kilogram and per day,
using the specific fiducial mass of each detector. 
The activities for $^{65}$Zn are calculated including the count rates observed in both
single- and multiple-hit events (see Section~\ref{sec-coinc})
and they are corrected for the effect of the 1115.5~keV gamma
to obtain total activities.
Monte Carlo simulations have shown that the probability for this gamma to escape
the fiducial volume without interacting is 50.1\%.
Given the 50.0\% branching ratio for the emission of this gamma, it can be estimated that
25\% of all $^{65}$Zn EC decays do not produce a visible 8.98 keV peak,
and the observed rate is corrected accordingly.

\section{Exposure history}
%%%%%%%%%%
\label{sec-history}

The population of cosmogenically produced radioactive isotopes in germanium 
is expected to be reduced to a negligible level during the crystallization process, 
except for $^{68}$Ge.
The activation occurs while the crystal is stored above ground, 
induced mostly by high-energy neutron interactions. 
After a time $\Delta t$ = $t$$-t_{g}$ after crystal growth happening at time $t_{g}$, 
the isotope activity per unit mass $a(t)$ increases as:

\begin{equation}
\label{eq:1}
a(t)  =  a_{max} \; (1-e^{-\frac{t-t_g}{\tau}})
\end{equation}

\noindent where $a_{max}$ is the maximal activation, 
obtained when it becomes equal to the production rate, 
and $\tau$ is the lifetime.
The value of $a_{max}$ for a given isotope depends on the flux of cosmic rays and 
 the cross section for its production (including decay chains).
Once the detector is shielded from cosmic rays by a significant thickness of rock
 so further activation can be neglected, the decay rate changes as:
\begin{equation}
\label{eq:2}
a(t)  =  a_{max} \; (1-e^{-\frac{t_s-t_g}{\tau}}) \; e^{-\frac{t-t_s}{\tau}}
\end{equation}

\noindent where $t_{s}$ is the time at which the detector is screened from cosmic rays.
A precise history detailing the succession and the length of the periods of time
during which the detector is exposed or protected from cosmic rays is
thus essential to evaluate isotope activities at a given time.

The crystals have been produced gradually from September 2011 to December 2012, and they have been installed at the LSM in three batches: in January 2013, in September 2013 and  in January 2014.
The time periods of exposure to cosmic rays and underground storage of the Ge detectors are listed per detector in Table~\ref{tab:history}.
The germanium crystals were first exposed above ground in a period $t_{exp1}$ 
starting from their production  by UMICORE in Belgium, 
during their polishing at BSI in Latvia, up to their shipment
to different shallow underground sites at CEA in France.
The shallow sites were at least 10~m.w.e. deep. 
Considering the attenuation length of the neutron flux of 148~g/cm$^{2}$~\cite{ziegler}, 
such depths result in a reduction of activation by a factor 850.  
Cosmogenic production during the time $t_{dec1}$ when the crystals were stored 
in the shallow sites is therefore neglected.
 The second exposure time $t_{exp2}$ occurs when the detector is shipped from 
 the shallow sites to the nearby laboratories of CSNSM and CEA, where electrodes are 
 evaporated, the surfaces are etched with XeF$_2$~\cite{leakage},
 the NTD heat sensors are glued to them,
 and the detectors are shipped to an underground site.
 This site is either the LSM, or, for three of the detectors,
 one of the shallow sites, where they were stored prior to a grouped shipment to LSM.
The last decay period $t_{dec2}$ corresponds to the interval between their
arrival at LSM and the date of October 30$^{th}$, 2014 ($t_{ref}$).
For the three detectors stored in the shallow site before their shipment to LSM,
this period is included in $t_{dec2}$, and the short shipment time is included
in the systematic uncertainty.

Confirmed dates are those of the crystal production, 
the end of their polishing, 
the start and end of the electrode evaporation, 
the XeF$_2$ etching processes,
the glueing of the NTD,
and of their arrival at the LSM underground laboratory. 
Tracking of detector shipments and in-and-out of the shallow sites 
are not documented as precisely. 
This is taken into account by ascribing a $\pm$7 day uncertainty on the duration of
each shipment, resulting in the systematic uncertainty shown in Table~\ref{tab:history}.
This uncertainty is anti-correlated between two consecutive periods.
The time periods $t_{exp1}$ and $t_{dec2}$ are bound by only one transport,
and therefore have a smaller uncertainty.
For two of the detectors, FID827 and FID828, there is a significantly larger uncertainty
on the time interval between detector fabrication and shipment to underground site.

Considering the succession of exposure times  ($t_{exp1}$ and $t_{exp2}$) and
cooling-off periods ($t_{dec1}$ and $t_{dec2}$), 
and using Eqs.~(\ref{eq:1}) and (\ref{eq:2})
the decay rate observed on October 30$^{th}$, 2014 ($t_{ref}$)
 in a detector ($\frac{dN}{dt}$) 
 is related to the production rate ($P$) by:
%%% 
\begin{equation}\label{eq:pr}
\frac{dN}{dt}   =  \nonumber\\
P \times [(1-e^{\frac{-t_{exp1}}{\tau}})(e^{\frac{-(t_{dec1}+t_{exp2}+t_{dec2})}{\tau}})  + (1-e^{\frac{-t_{exp2}}{\tau}})(e^{\frac{-t_{dec2}}{\tau}})]
\end{equation}

The expression in brackets in Eq.~(\ref{eq:pr}) corresponds to the saturation fraction $f_{s}(t_{ref})$,  i.e.\
the activation at the time $t_{ref}$ expressed as a fraction of $a_{max}$, for a given detector history.
Table~\ref{tab:saturationfraction} lists $f_{s}$ values calculated from Table~\ref{tab:history},
together with their systematic uncertainties $\sigma(f_{s})$.

\begin{figure}
\centering
\begin{tabular}{lll}\\
\includegraphics[scale=0.3]{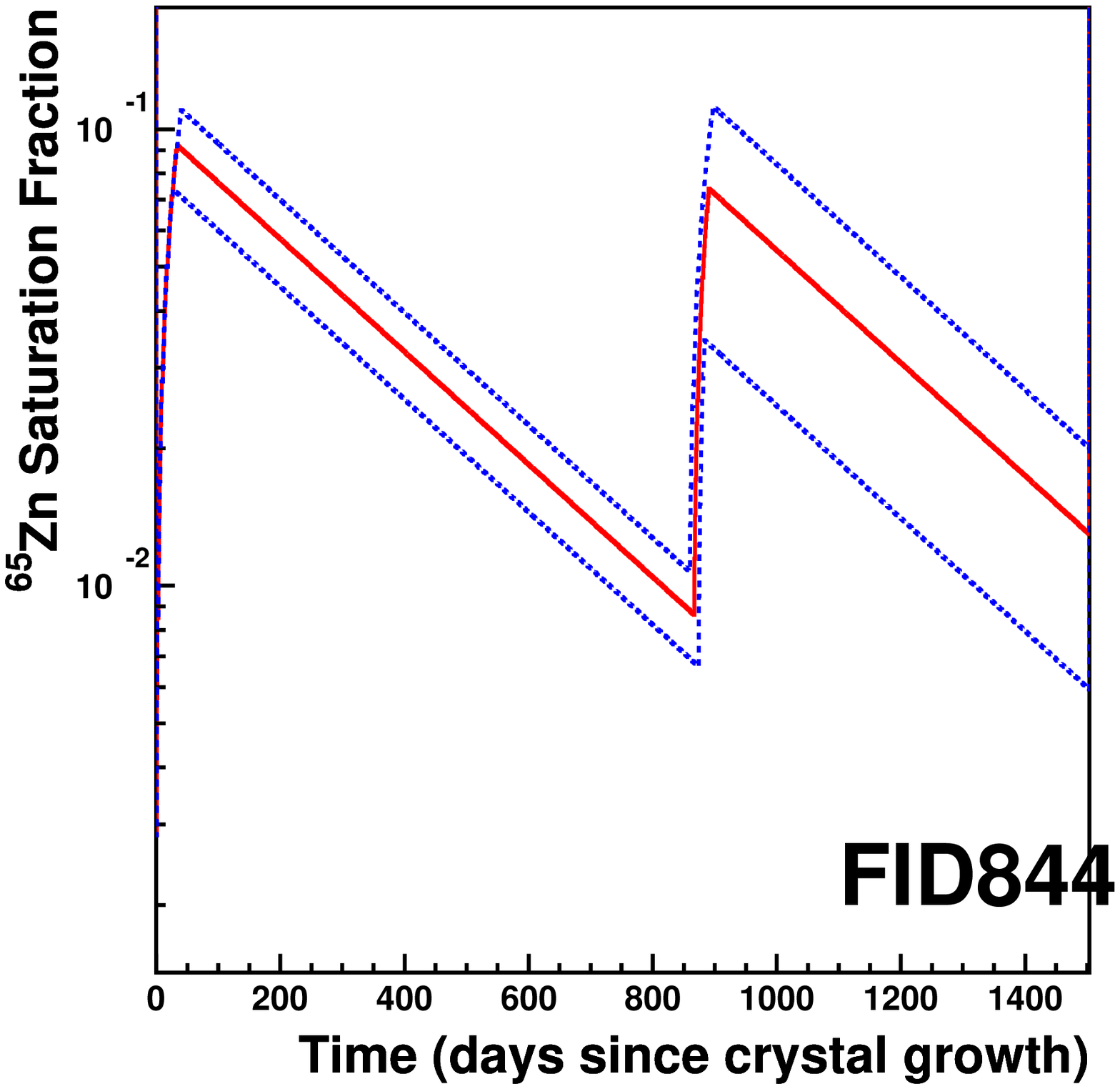}& 
 \includegraphics[scale=0.3]{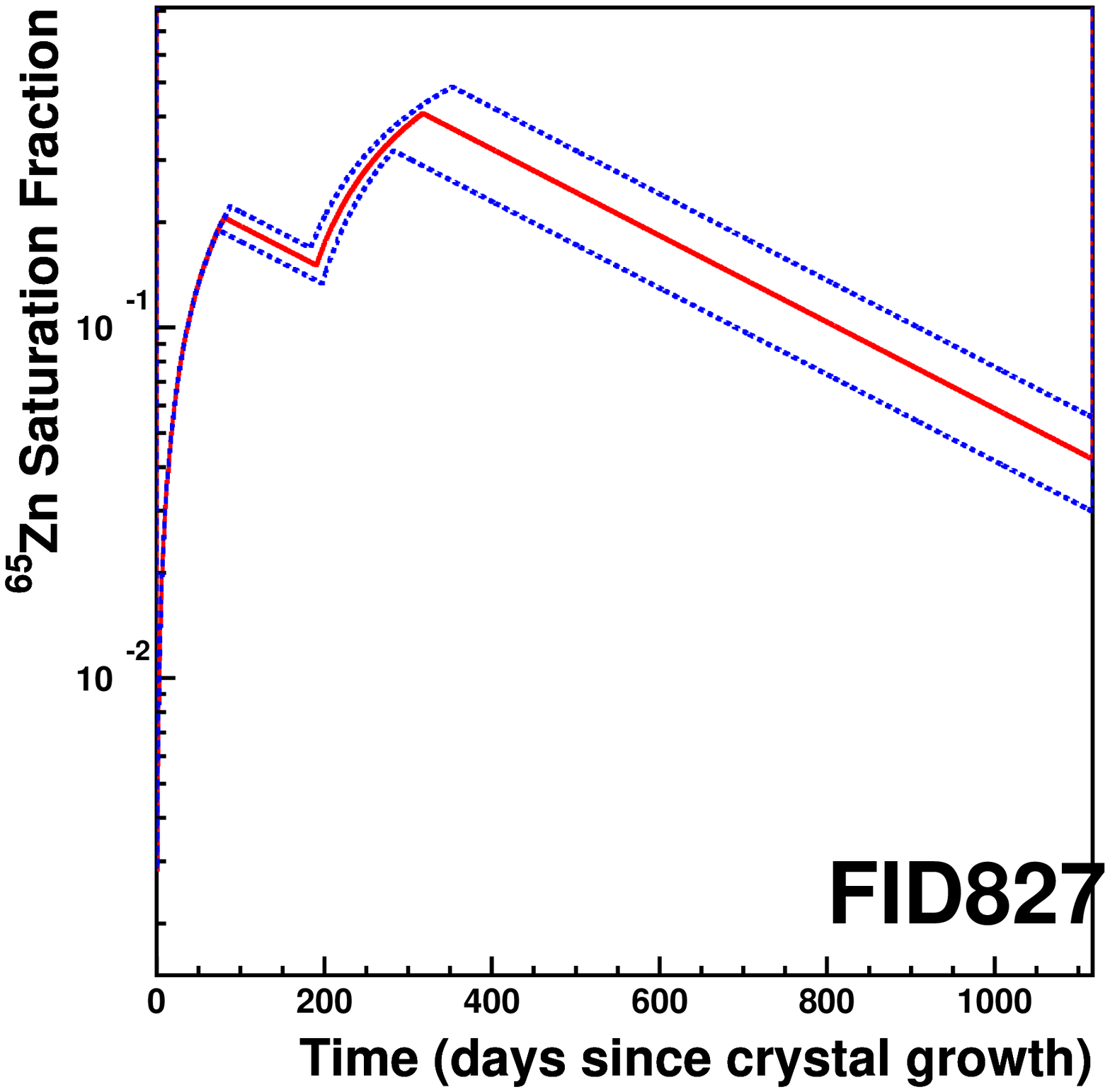}  \\
\includegraphics[scale=0.3]{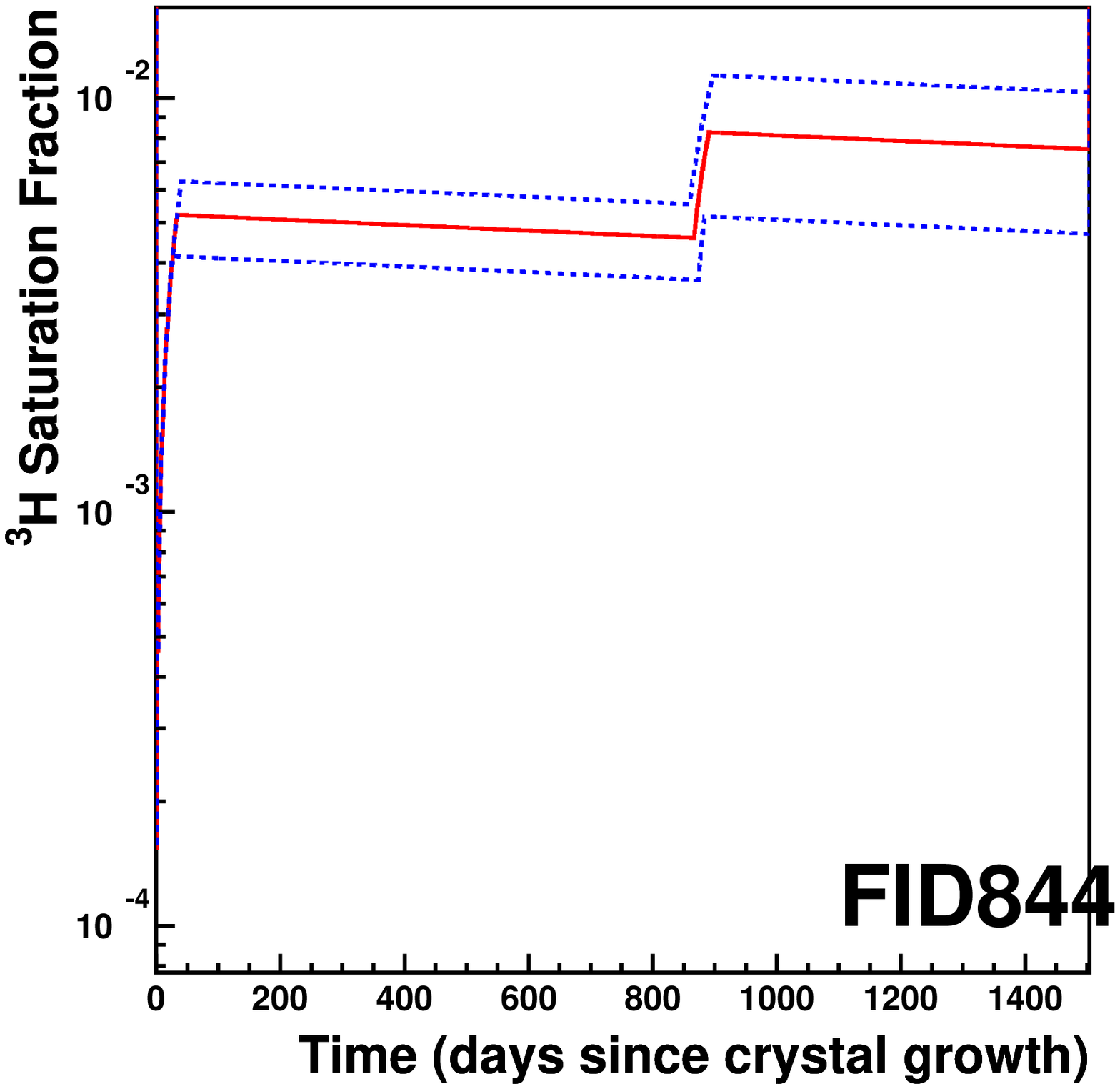}& 
 \includegraphics[scale=0.3]{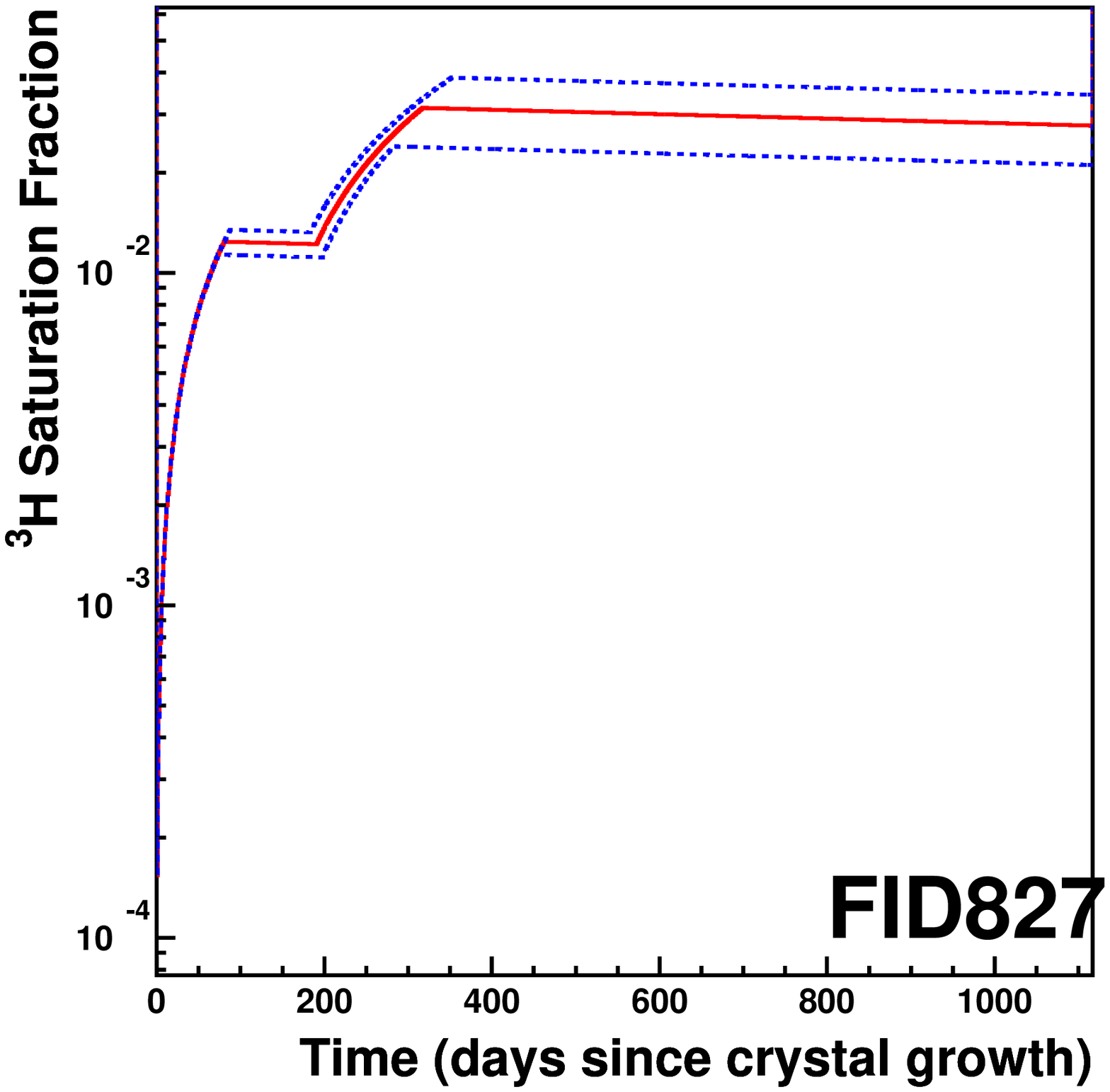}  \\
\end{tabular}
\caption{Saturation fraction $f_{s}$ as a function of time for detector FID844 (standard history, left figure) and FID827 (longer $t_{exp2}$, right figure). Red plain lines represent the saturation fraction for the average  $t_{exp1}$, $t_{exp2}$, $t_{dec1}$ and $t_{dec2}$, whereas the blue bands represent the systematic uncertainty due to the history model considered. Top and bottom rows refer to  $^{65}$Zn and $^{3}$H, respectively. 
\label{fig:satfrac}}

\end{figure}

The calculation of $f_s$ can also be done at any other time $t$ between $t_g$ and $t_{ref}$. This is what
is shown on Fig.~\ref{fig:satfrac} for $^{65}$Zn (top) and $^{3}$H (bottom),
for the two detectors FID844 (left) and FID827 (right).
The time axis starts at crystallization and ends on the reference day $t_{ref}$.
The red lines correspond to the history described in Table~\ref{tab:history},
while the blue dotted lines are calculated using the $\pm$1$\sigma$ uncertainties.
The detector FID844 has a standard history, where the two
exposures $t_{exp1}$ and $t_{exp2}$ are separated by a relatively long
cooling-down period $t_{dec1}$.
This period is sufficiently long compared to the mean lifetime of $^{65}$Zn
that  the activity at $t_{ref}$ is dominated by the effect of the last exposure $t_{exp2}$,
and consequently
the uncertainty on $f_{s}(t_{ref})$ 
is dominated by that on $t_{exp2}$.
This is not the case for $^{3}$H ($t_{dec1}\ll \tau$), where both exposure times
$t_{exp1}$ and $t_{exp2}$ contribute equally to $f_{s}(t_{ref})$.
FID827 is an example of a different history, where $t_{dec1}$ is also
small compared to the $^{65}$Zn lifetime.

As a first consistency test of the detector history model with the observed count rates,
the correlation factor between the measured rate and the saturation fraction of $^{68}$Ge and $^{65}$Zn, 
for which measurements provide higher statistics,  were evaluated.
The correlation factors, calculated as Pearson product-moment correlation coefficient, 
are 0.90 and 0.83, respectively. 
This is a good indication that the observed count rate variations from detector to detector are 
indeed caused by the known history differences between them.
In the final analysis (next section), the validity of this correlation is tested taking into account
the uncertainties of the measurements and detector history.

\section{Production rates}\label{sec:pr}

%%%%%%%%%%%
\subsection{Experimental Results}

Figures~\ref{fig:tritium} to \ref{fig:ge68} show the
measured decay rates of Table~\ref{tab:rates}, 
as a function of the saturation fraction of Table~\ref{tab:saturationfraction}, 
for the thirteen FID detectors. 
Each plot corresponds to a different isotope.
The data are fitted with a first degree polynomial: $y = P\cdot~x$, where $x$ is the saturation fraction, $y$ is the measured decay rate and the coefficient $P$ is the production rate of this particular isotope. 
The fits are shown by solid red lines.
Errors in the count rates and saturation fraction have been taken into account.
The fit parameters $P$ corresponding to the best estimates of the production rate are listed in Table~\ref{tab:pr}, together with the $\chi^2/d.o.f.$ of the fit.
The final production rate values and their total errors are listed in Table~\ref{tab:results}. 

Most fits result in good reduced $\chi^2$ values, except $^{65}$Zn and $^{3}$H. 
To investigate this further, 
the fits were repeated with different scaling factors $S$ applied to the systematic uncertainty in the saturation fraction. 
The reduced $\chi^2$ values for $^{65}$Zn and $^{3}$H are equal or less than one for $S=2$.
We therefore include as an additional systematic uncertainty in the production rate of 
these two isotopes  the difference between the results with $S$ = 1 and $S$ = 2.
The final production rate values and their total errors are listed in Table~\ref{tab:results}.

Additional checks for consistency of the results have been carried out.
Three detectors (FID826, 827 and 828) have been removed from the analysis because they were stored at a shallow site after being fully assembled but before being 
shipped to LSM, which could induce additional uncertainty relative to other detectors that
were shipped directly to LSM.
Production rates per isotopes agree within statistical errors with initial results.
Although no activation is assumed to occur at the shallow sites,
we repeated the analysis considering an activation rate equivalent to either  2\% or 5\% 
of the value on the surface.
This resulted in small changes in the saturation fraction at $t=t_{ref}$ which are within the systematic uncertainty arising from the shipment history alone.

\subsection{Calculated production rates}

\begin{figure}
\begin{center}
 \includegraphics[scale=0.55]{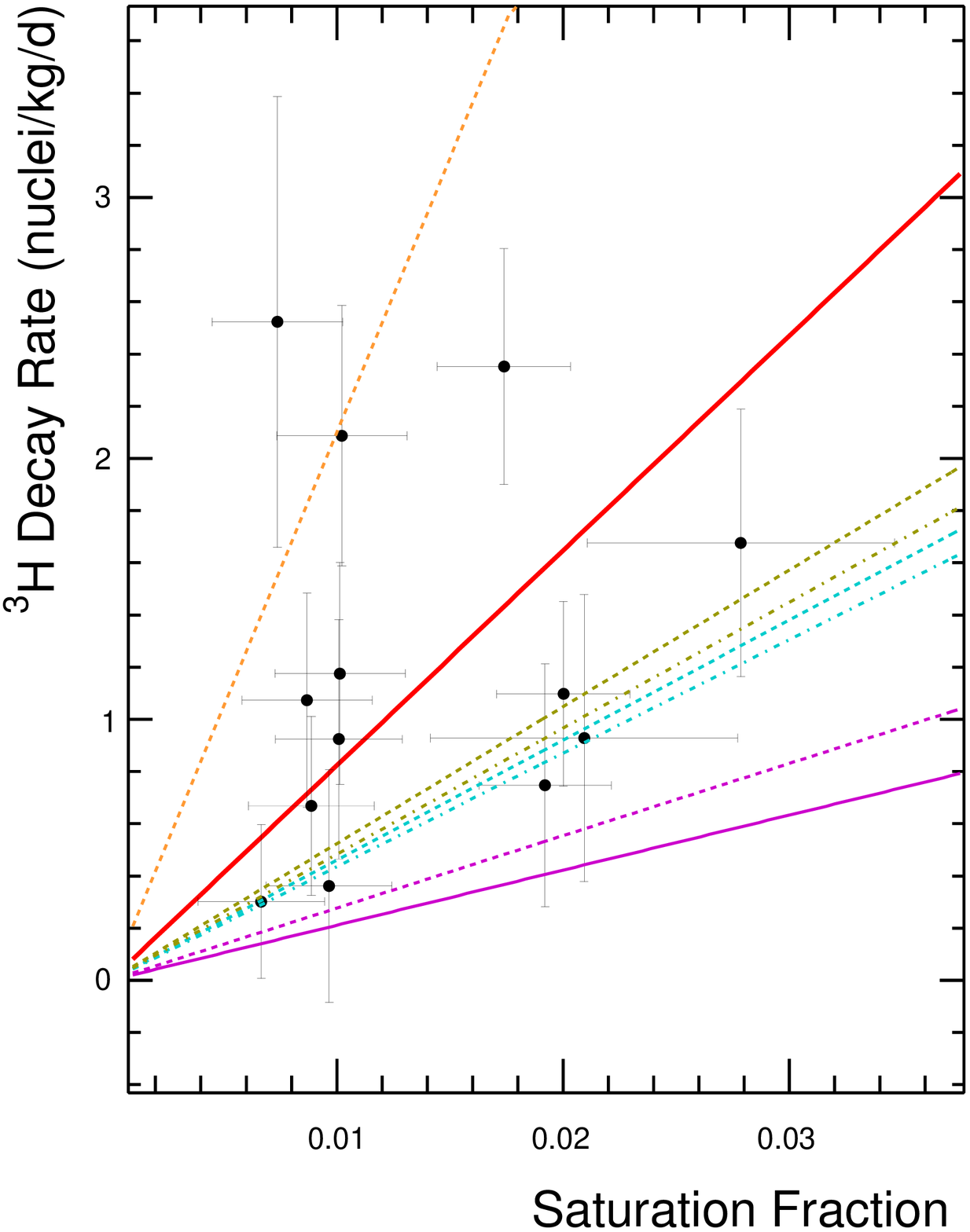} \end{center}
\caption{Tritium decay rate as a function of saturation fraction for thirteen detectors. Horizontal error bars show uncertainties in exposure and decay time periods. Vertical error bars include statistical and systematic errors. Red solid line represents the best fit. ACTIVIA~\cite{activia} calculations performed within this work are shown as dashed and dashed-dotted light-blue lines using semi-empirical~\cite{silb,silb1,silb2,silb3,silb4} and MENDL-2P~\cite{mendl} cross sections, respectively. Purple solid line represents an upper limit from IGEX data estimates in Ref.~\cite{mei}. Previous calculations have been included as dashed lines:  Mei et al.~\cite{mei} using TALYS cross sections (purple), an update of this latter from Zhang et al.~\cite{Zhang:2016rlz} using GEANT4 (olive-green dashed line) and ACTIVIA (olive-green dotted-dashed line) and Avignone et al.~\cite{avignone} from~\cite{Hess} (orange). 
\label{fig:tritium}}

\end{figure}

\begin{figure}
\begin{center} 
\includegraphics[scale=0.55]{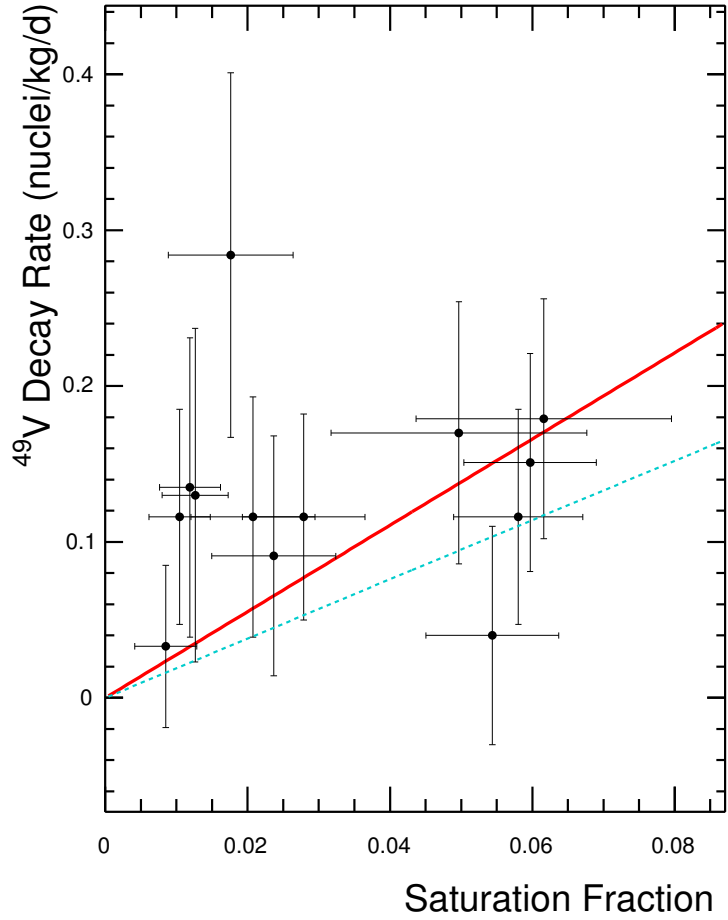} \end{center}
\caption{$^{49}$V decay rate as a function of saturation fraction for thirteen detectors. Horizontal error bars show uncertainties in exposure and decay time periods. Vertical error bars include statistical and systematic errors. Red solid line represents the best fit. ACTIVIA~\cite{activia} calculations using semi-empirical~\cite{silb,silb1,silb2,silb3,silb4} and MENDL-2P~\cite{mendl} cross sections performed within this work give the same result that is shown as dashed light-blue line.
\label{fig:v}}
\end{figure}

\begin{figure}
\begin{center} 
\includegraphics[scale=0.55]{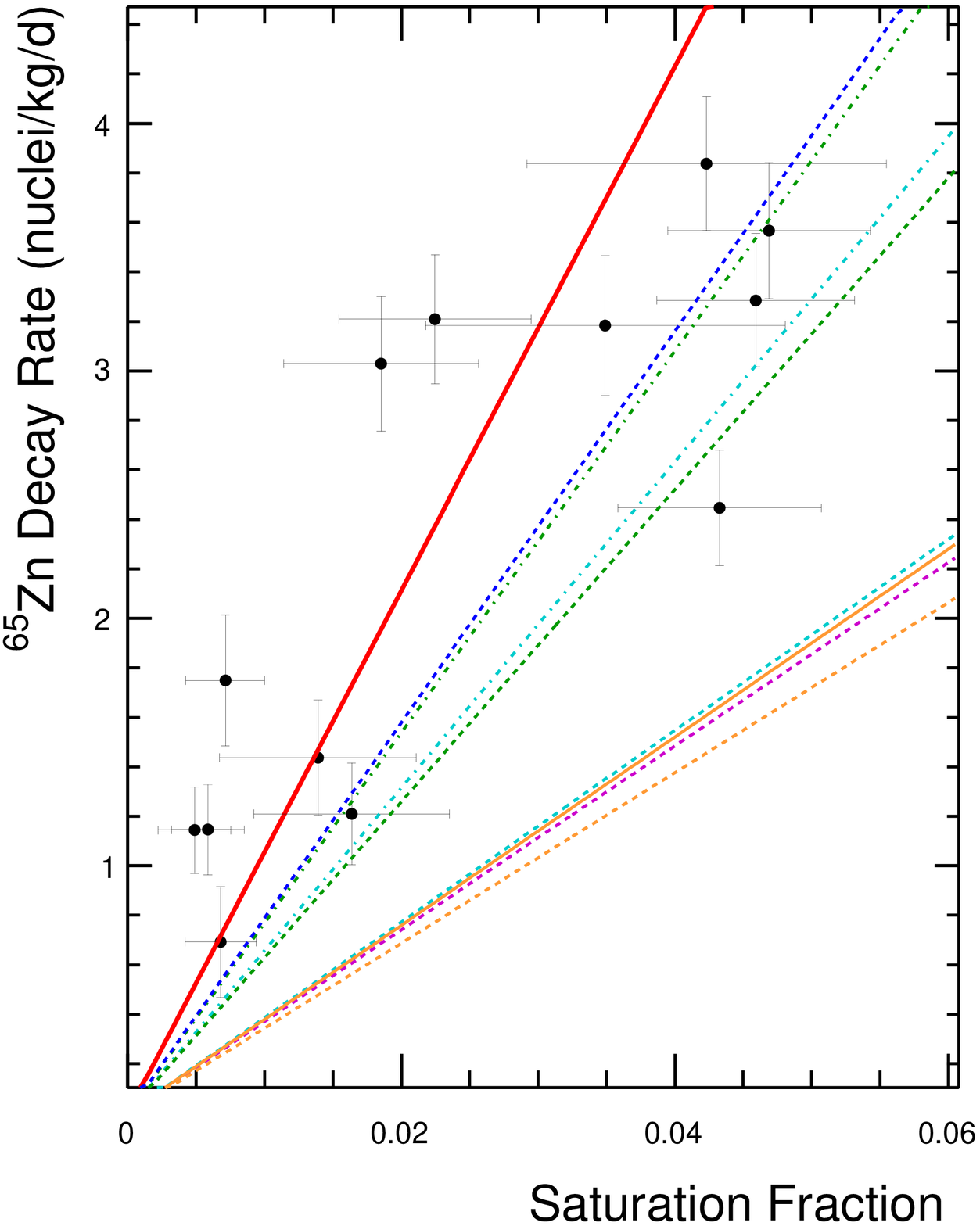} \end{center}
\caption{$^{65}$Zn decay rate as a function of saturation fraction for thirteen detectors. Horizontal error bars show uncertainties in exposure and decay time periods. Vertical error bars include statistical and systematic errors. Red solid line represents the best fit. ACTIVIA~\cite{activia} calculations performed within this work are shown as dashed and dashed-dotted light-blue lines using semi-empirical~\cite{silb,silb1,silb2,silb3,silb4} and MENDL-2P~\cite{mendl} cross sections, respectively.  
Previous calculations have been included as dashed lines: Klapdor-Kleingrothaus et al.~\cite{KK} (dark blue), Mei et al.~\cite{mei} using TALYS cross sections (purple), Avignone et al.~\cite{avignone} (orange). The calculations from Ref.~\cite{ceb} that assume Ziegler~\cite{ziegler} and Gordon et al.~\cite{gordon} cosmic neutron spectra are shown as green dotted and dashed-dotted lines, respectively. In addition, solid orange line shows previous experimental data from Ref.~\cite{avignone}. 
\label{fig:zn}}
\end{figure}

\begin{figure}
\begin{center} 
\includegraphics[scale=0.55]{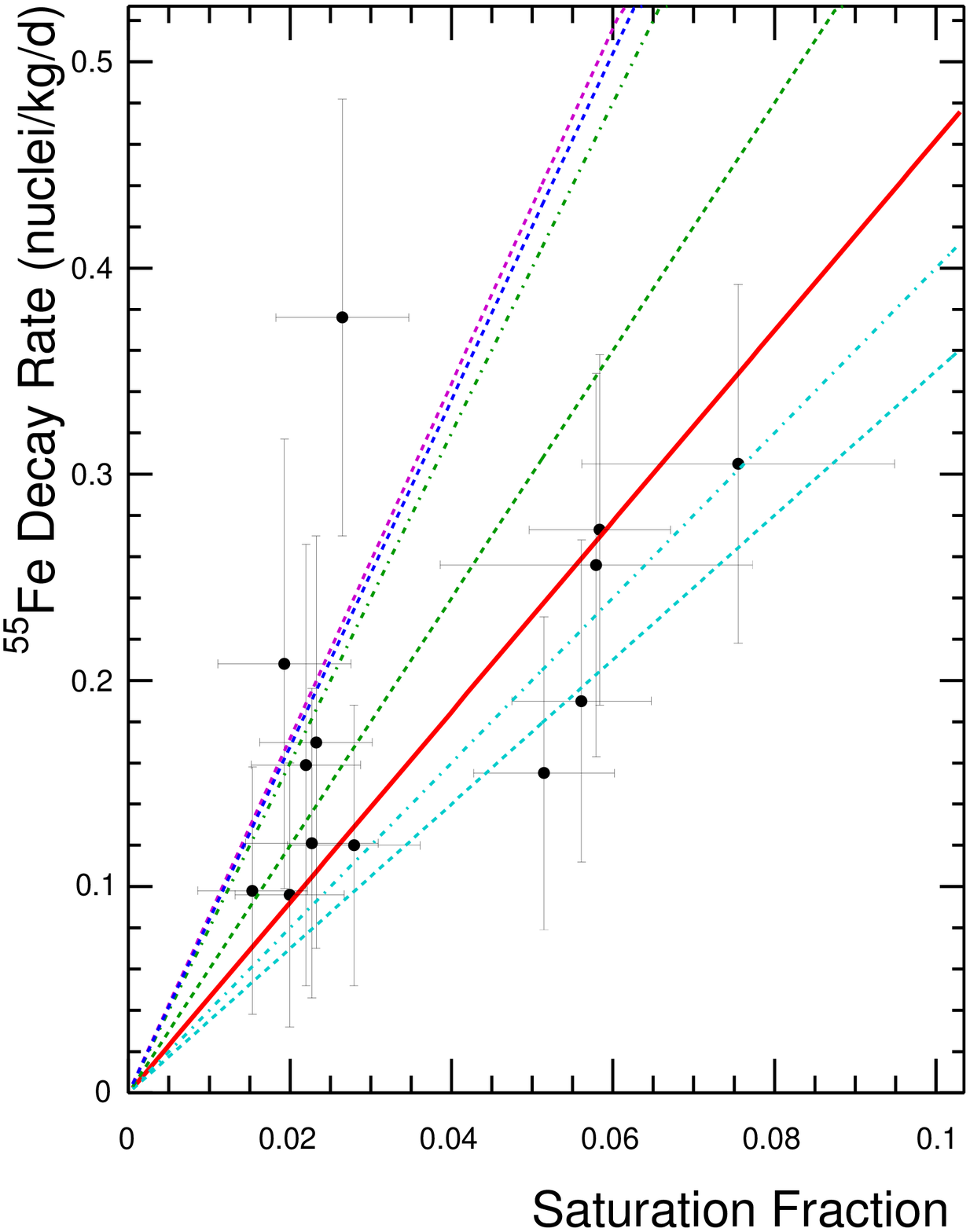} \end{center}
\caption{$^{55}$Fe decay rate as a function of saturation fraction for thirteen detectors. Horizontal error bars show uncertainties in exposure and decay time periods. Vertical error bars include statistical and systematic errors. Red solid line represents the best fit. ACTIVIA~\cite{activia} calculations performed within this work are shown as dashed and dashed-dotted light-blue lines using semi-empirical~\cite{silb,silb1,silb2,silb3,silb4} and MENDL-2P~\cite{mendl} cross sections, respectively. 
Previous calculations have been included as dashed lines: Klapdor-Kleingrothaus et al.~\cite{KK} (dark blue), Mei et al.~\cite{mei} using TALYS cross sections (purple). The calculations from Ref.~\cite{ceb} that assume Ziegler~\cite{ziegler} and Gordon et al.~\cite{gordon} cosmic neutron spectra are shown as green dotted and dashed-dotted lines, respectively.
\label{fig:fe}}
\end{figure}

\begin{figure}
\centering
\begin{tabular}{ll}\\
\includegraphics[scale=0.35]{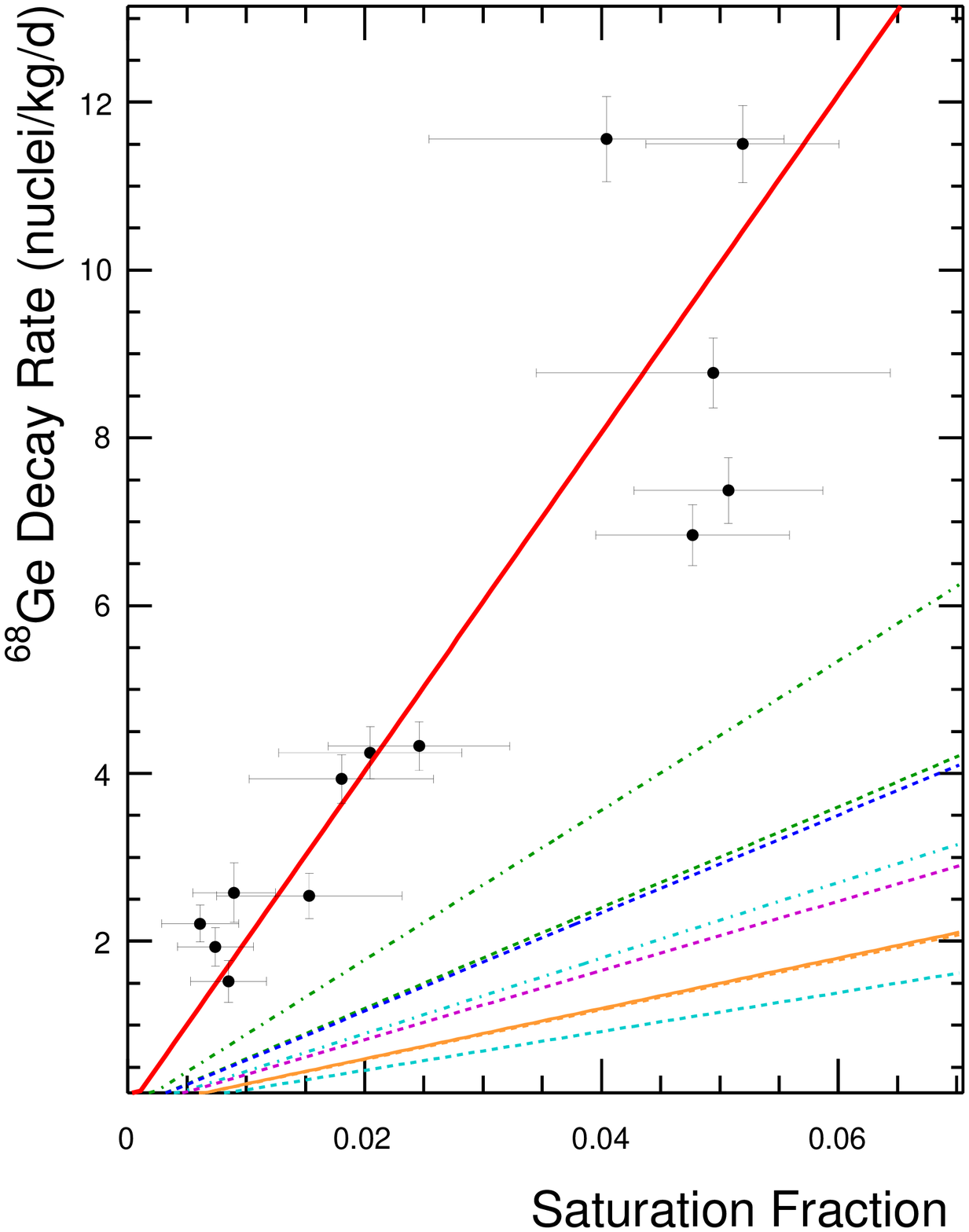} &  \includegraphics[scale=0.35]{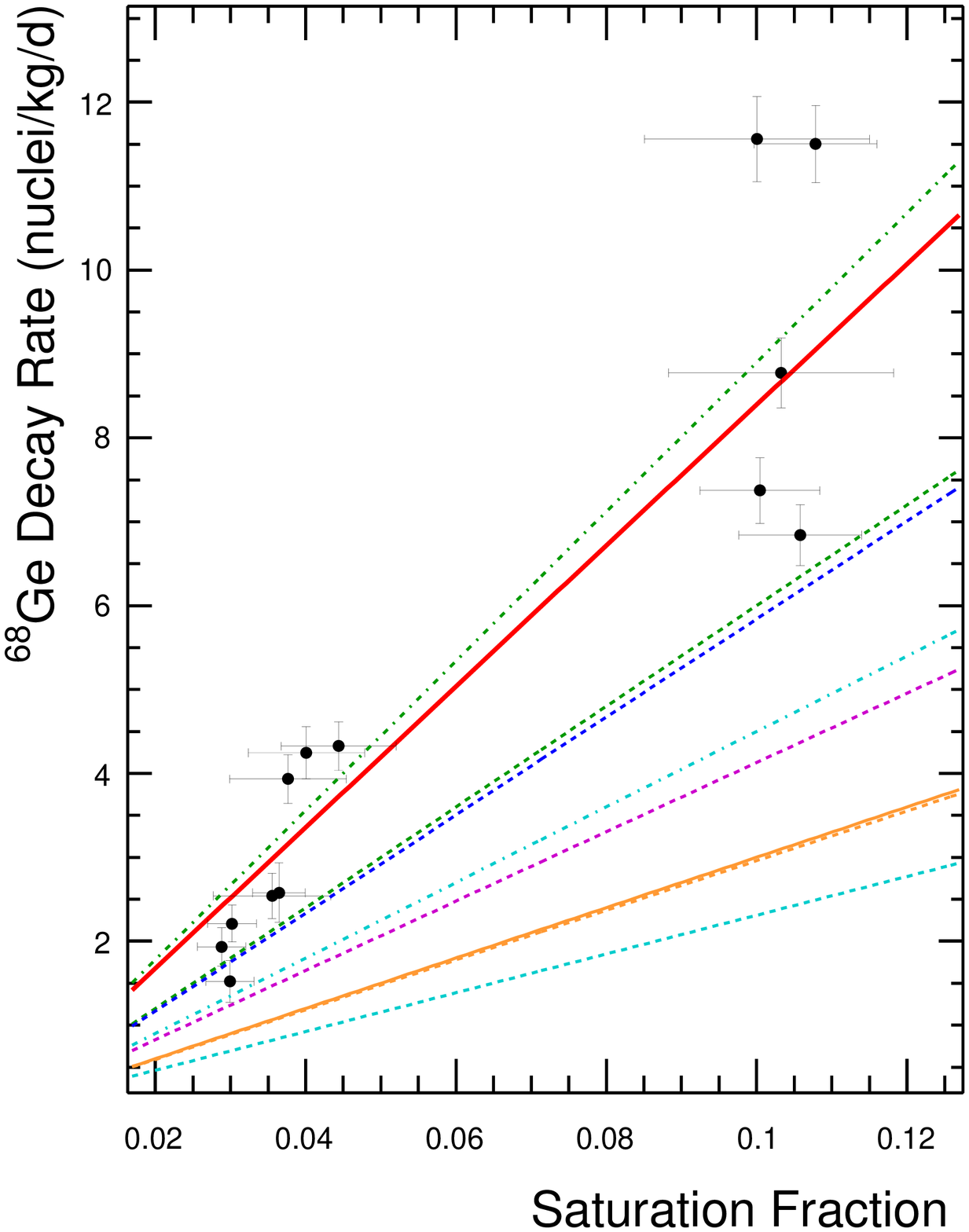} 
\end{tabular}\\
\caption{$^{68}$Ge  decay rate as a function of saturation fraction for thirteen detectors. Horizontal error bars show uncertainties in exposure and decay time periods. Vertical error bars include statistical and systematic errors. Red solid line represents the best fit. ACTIVIA~\cite{activia} calculations performed within this work are shown as dashed and dashed-dotted light-blue lines using semi-empirical~\cite{silb,silb1,silb2,silb3,silb4} and MENDL-2P~\cite{mendl} cross sections, respectively. 
Previous calculations have been included as dashed lines: Klapdor-Kleingrothaus et al.~\cite{KK} (dark blue), Mei et al.~\cite{mei} using TALYS cross sections (purple), Avignone et al.~\cite{avignone} (orange). The calculations from Ref.~\cite{ceb} that assume Ziegler~\cite{ziegler} and Gordon et al.~\cite{gordon} cosmic neutron spectra are shown as green dotted and dashed-dotted lines, respectively. In addition, solid orange line reflects previous experimental data in Ref.~\cite{avignone}.  Left panel represents the saturation fraction and production rate fit considering the standard detector history, whereas the right panel considers an additional 3-year exposure before the crystallization process. It reflects the saturation of $^{68}$Ge. \label{fig:ge68}}
\end{figure}

Figures~\ref{fig:tritium} to \ref{fig:ge68} also show the lines corresponding to 
production rate estimates from different models.

The production rate, $P_{i}$, of a radioactive isotope $i$ can be calculated as follows:
\begin{equation}\label{eq:pr1}
P_{i} = \sum_{j} N_{j} \int\phi(E)\sigma_{ij}(E) dE
\end{equation}
 where $N_{j}$ is the number of target nuclear isotope $j$, 
 $\sigma _{ij}$ is the excitation function of isotope $i$ produced  by neutrons on stable isotopes of material $j$, 
 and $\phi$ is the cosmic neutron flux. 
 
Estimates of the production rate have been carried out with the ACTIVIA code~\cite{activia}. The code does not include proton activation of materials; the hadronic part of the cosmic-ray spectrum at the surface is dominated by neutrons~\cite{ziegler}.
The original ACTIVIA code uses the neutron energy spectrum at the surface from~\cite{ziegler} but, as shown in Ref.~\cite{Zhang:2016rlz}, this spectrum does not match recent experimental data~\cite{gordon}. We carried out the calculations of production rates by replacing the original neutron spectrum with that from Ref.~\cite{gordon}.

These calculations are performed using  a cosmic ray neutron spectrum in the northern hemisphere
parametrized as follows~\cite{gordon}:
\begin{eqnarray}\nonumber\label{eq:nflux}
\phi(E)  &=& 1.006 \times 10^{-6}e^{-0.35\ln^{2}E+2.1451\ln{E} } \\
&& + 1.011 \times 10^{-3}e^{-0.4106\ln^{2}E-0.667\ln{E}} 
\end{eqnarray}
where $E$ is the neutron kinetic energy in MeV 
and $\phi$ is in units of cm$^{-2}$~s$^{-1}$~MeV$^{-1}$. 

We have also used semi-empirical cross sections from Refs.~\cite{silb,silb1,silb2,silb3,silb4} and the cross sections from the MENDL-2P libraries~\cite{mendl}, shown, respectively, as dashed and dash-dotted light-blue lines in Figs.~\ref{fig:tritium} to~\ref{fig:ge68}.

The results of the calculations are listed in the third column of Table~\ref{tab:results}. 
They are labeled (a) or (b) if they use the 
semi-empirical cross sections~\cite{silb,silb1,silb2,silb3,silb4} or the MENDL-2P~\cite{mendl} database, respectively. 
For  $^{49}$V both calculations give the same production rate.
Our calculations with ACTIVIA, quoted in the second column of Table~\ref{tab:results}, show that excitation functions may account for up to a factor of 2 difference in the production rate of $^{68}$Ge. The difference increases with the atomic number of the isotope produced.

Effects on the production rates caused by differences in cross sections 
and cosmogenic neutron energy spectra in ACTIVIA and GEANT4 codes~\cite{geant} 
have been estimated in Ref.~\cite{Zhang:2016rlz}.  
There, the neutron spectrum from~\cite{gordon} was taken as measured values for New York location whereas in the present work the parameterization~\cite{gordon} of these data has been used as given by Eq.~(\ref{eq:nflux}). The two approaches (data vs parameterization of these data) differ by no more than 2\%. 

Comparing production rates from calculations in Ref.~\cite{ceb}, listed in the third and fourth columns of Table~\ref{tab:results}, we can conclude that different input cosmic-ray neutron spectra, from Refs.~\cite{ziegler} and \cite{gordon},  can lead to a variation in production rates of about 20-30\%  for the isotopes considered here.

\subsection{Comparison with previous measurements and models}

Different model estimates have been compared to the measurements presented in the previous sections. 
In  Figs.~\ref{fig:tritium} to~\ref{fig:ge68}, the EDELWEISS data and the best fit to these data are compared to
the ACTIVIA calculations performed within this work
considering semi-empirical~\cite{silb,silb1,silb2,silb3,silb4} (dashed light-blue line) and MENDL-2P~\cite{mendl} (dash-dotted light-blue line) cross sections. 
Calculations from previous works are also shown in dashed lines:
 Cebrian et al.~\cite{ceb} (green), 
 Klapdor-Kleingrothaus et al.~\cite{KK} (blue), 
 and Mei et al.~\cite{mei} using TALYS cross sections (purple), 
 the update of this work  from Zhang et al.~\cite{Zhang:2016rlz} (olive-green)
and the estimates from Avignone et al.~\cite{avignone} (orange). 

The calculations from Ref.~\cite{ceb} that assume Ziegler~\cite{ziegler} and Gordon et al.~\cite{gordon} 
cosmic neutron spectra
are shown as dotted and dashed-dotted lines, respectively.  
All these calculated rates are also listed in Table~\ref{tab:results}.

In addition, the previous experimental measurements for $^{68}$Ge and $^{65}$Zn reported in Ref.~\cite{avignone}
are shown as solid orange line in  Fig.~\ref{fig:zn} and Fig.~\ref{fig:ge68} and appear in the last column of
Table~\ref{tab:results}.

Among themselves, 
these models agree within a factor of 2-3 for $^{65}$Zn, for $^{68}$Ge and $^{55}$Fe, 
whereas the difference for $^{3}$H can be close to one order of magnitude, 
ranging from 27.7 to 210 nuclei/kg/day.

Ref.~\cite{mei} quotes an experimental upper limit on the $^{3}$H production rate of 21~nuclei/kg/day of $^{3}$H from a by-eye fit to the IGEX data, 
in tension with an estimate of 27.7~nuclei/kg/day from the model using the TALYS cross sections 
and the cosmic neutron spectrum parameterization from Ref.~\cite{gordon}.  
A comparison of the latter value with most recent calculations is available in Ref.~\cite{Zhang:2016rlz}. 
However, the tritium production rates estimated with GEANT4 (I) and ACTIVIA (II) are larger: 
52.4~nuclei/kg/day  and 48.3~nuclei/kg/day, respectively, 
and are closer to our measurement of  82$\pm$21~nuclei/kg/day with statistical and systematic uncertainties included. 

The measured production rates for all nuclei are within a factor two of our ACTIVIA calculations, 
with preferred cross sections from MENDL-2P library, especially for the
nuclei $^{65}$Zn and $^{68}$Ge.
The closest agreement between data and our ACTIVIA calculations is for $^{55}$Fe.
Concerning the other calculations listed in Table~\ref{tab:results}, those
predicting the largest rates are systematically favored,
except for the $^{3}$H prediction from Ref.~\cite{avignone}.

The significance of the tension between our measurements and ACTIVIA calculations
can be assessed by comparing the difference in their values to the quoted
experimental uncertainties. 
The ACTIVIA estimates from this work show a  1.7$\sigma$ to 1.8$\sigma$
deviation from the measured production rate  of $^{3}$H of 82~$\pm$~21~nuclei/kg/day.  
The disagreement with the prediction from Ref.~\cite{avignone} using the prescription from \cite{Hess} is stronger, at 6$\sigma$.
The agreement between the measured rates and those predicted by ACTIVIA (average value), 
for isotopes other than $^{3}$H, is better:
deviations of less than 1.7$\sigma$ (statistical error only) are observed for $^{55}$Fe, $^{49}$V and $^{65}$Zn. 
%%%%%%
It should be noted that the present measurement for $^{65}$Zn disagrees with that  
in Ref.~\cite{avignone} by 5.9$\sigma$ (statistical error only). 
Including systematic uncertainty, the disagreement is still strong at a 4.8$\sigma$ level.

Since the population of $^{68}$Ge is not reduced  to zero during the germanium crystallization, 
there is an uncertainty in its saturation fraction
associated to the amount of time the germanium ore was exposed to cosmic rays before the crystal growing.
The first value for $^{68}$Ge in Table~\ref{tab:pr}  (202 $\pm$ 16)  is calculated assuming no exposure prior
to the crystallization.
The second assumes a 3-year exposure before the crystal growth, reaching
a population close to saturation.
In this second scenario, the production rate is reduced to 84~$\pm$~3~nuclei/kg/day. The $\chi^{2}$ of the fit is reduced from 3.2 to 1.0 assuming a scaling factor of $\sigma(f_s)$ of 2 (see Section 6.1)
resulting in a production rate of 84 $\pm$ 6, 
from which is obtained a lower limit on the production rate of 74~nuclei/kg/day at 90\% C.L.\.
It is, nevertheless,  3.6$\sigma$ above the average value of the ACTIVA results, considering the difference between the two predictions
as a systematic uncertainty.
Ref.~\cite{avignone} reports a value of 30~$\pm$~7~nuclei/kg/day at saturation, 
6.3$\sigma$ below our lower limit. 
The present $^{65}$Zn production rate measurement is 2.8~$\pm$~0.6 times larger than that of Ref.~\cite{avignone}. 
The lower limit on the $^{68}$Ge value is also larger than the measurement in that same reference, by a similar factor of 2.5.
In order to evaluate the possible source of this discrepancy, more detailed information on the analysis of the data of Ref.~\cite{avignone} are needed. 

Our measurements, extending to more than one cosmogenic activation product, 
can help to improve the models by constraining them better and thus contribute to
the reduction of
the systematic uncertainties associated with the wide variation of their predictions.
The discrepancies of predictions from nucleus to nucleus, and the significant difference in the ACTIVIA calculations
with semi-empirical or MENDL-2P cross sections suggest that an important source of uncertainties in the calculations comes from the different excitation functions of an isotope produced by neutrons on the stable  parent isotope of the material.

\section{Conclusion}

The cosmogenic activation of various isotopes in the 
germanium detectors of the  EDELWEISS-III  experiment has been measured.
The data for five isotopes and thirteen detectors with different exposure times
lead to a consistent set of measurements.
The first measurement of the $^{3}$H decay rate in germanium detectors is presented. 
It has been interpreted in terms of production rate of 82~$\pm$~21~nuclei/kg/day  with statistical and systematic uncertainties included. 
The tritium production due to cosmic-ray neutrons is thus important and the
present measurement provides valuable information needed to evaluate
the reduction of the exposure to cosmic rays necessary for 
germanium detector arrays used for dark matter searches.
The measured production rates on $^{49}$V, $^{55}$Fe and $^{65}$Zn of 2.8 $\pm$ 0.6~nuclei/kg/day,
4.6 $\pm$ 0.7~nuclei/kg/day and
106 $\pm$ 13~nuclei/kg/day, respectively, 
presented here are the most accurate to-date.
A lower limit of 74~nuclei/kg/day at 90\% C.L.\ on production rate of $^{68}$Ge is discussed.

The measured $^{65}$Zn production rate and the lower limit on that of $^{68}$Ge are 
a factor 2.7 $\pm$ 0.6 larger than the measurements reported in Ref.~\cite{avignone}. 
The origin of this discrepancy is unknown.

The measurements agree within a factor of two with estimates performed with the ACTIVIA 
code within this work. The best agreement is found for $^{49}$V and $^{55}$Fe.
The estimates for $^{3}$H, $^{65}$Zn and $^{68}$Ge tend to underestimate the measured 
rates, with significance ranging from 1.7$\sigma$ to 6$\sigma$. 
The difference between these predictions and those from other models can also 
differ by as much as a factor of two in most cases, 
with no single model giving a satisfying description for all measured isotopes. 
It can be foreseen that the precision of the present measurements will help
constrain and further improve the models.

\section{Acknowledgments}

The help of the technical staff of the Laboratoire Souterrain de Modane and the participant 
laboratories is gratefully acknowledged. The EDELWEISS project is supported in part by the 
German ministry of science and education (BMBF Verbundforschung ATP Proj.-Nr.\ 05A14VKA), 
by the Helmholtz Alliance for Astroparticle Physics (HAP), by the French Agence Nationale 
pour la Recherche (ANR) and the LabEx Lyon Institute of Origins (ANR-10-LABX-0066) of 
the Universit{\'e} de Lyon within the program ``Investissements d'Avenir'' (ANR-11-IDEX-00007), 
by the P2IO LabEx (ANR-10-LABX-0038) in the framework "Investissements d'Avenir'' 
(ANR\-11\-IDEX\-0003\-01) managed by the ANR (France), by Science and Technology Facilities 
Council (UK), and the Russian Foundation for Basic Research (grant No. 15-02-03561).

\clearpage
\newpage
\nocite{*}
\bibliographystyle{aipnum-cp}
\bibliography{tritium}

\begin{table}\caption{Table of electron capture (EC) decays with half-lives above 10 days for 21$\leq$Z$\leq$33. 
The half-lives and decay properties are from Ref.~\cite{nudat}. 
The binding energy of electrons  in the $K$-shell ($E_K$)  are from Ref.~\cite{binding}, 
as well as those for $L$-shell electrons that are used in the fit shown in the inset of Fig.~\protect{\ref{fig-spectrum}}.
\label{tab-decay}}
\begin{center}
\begin{tabular}{c|c|c|c|c|c|c}
 \toprule
                  & Daughter    & $E_K$ &$E_L$ &   Half-life   & EC B.R. & Coincidences with $\gamma$ rays  \\
                   &    Isotope                  & (keV)    &(keV)    & $t_{1/2}$ & (\%)     &   and internal conversion   \\

\hline 
$^{74}$As & $^{74}$Ge & 11.10  & &  17.8 d  & 66           &\\
\hline
$^{73}$As & $^{73}$Ge & 11.10  & &  80.3 d  & 100         & \\
\hline
$^{71}$Ge & $^{71}$Ga & 10.37  & 1.30 &   11.4 d  & 100         & \\
\hline
$^{68}$Ge & $^{68}$Ga & 10.37  & 1.30 & 271.0 d  & 100         & \\
$^{68}$Ga & $^{68}$Zn &    9.66  & 1.19 & 67.7 min  & 11.12         & \\
\hline
$^{65}$Zn & $^{65}$Cu &    8.98  & 1.10& 243.9 d & 100         & 50.0\% (1115.5 keV) \\
%\hline
%$^{56}$Ni & $^{56}$Co &    7.71  &&    6.1 d & 100         & 100\% \\
\hline
$^{58}$Co & $^{58}$Fe &    7.11  &&   70.9 d & 100         & 99.5\% (810.8 keV)\\
\hline
$^{57}$Co & $^{57}$Fe &    7.11  &&  271.7 d & 100        & 99.8\% (122.1 or 136.5 keV)\\
\hline
$^{56}$Co & $^{56}$Fe &    7.11  &&   77.2 d & 100         & 100\% (846.8 keV and others)\\
\hline
$^{55}$Fe & $^{55}$Mn&    6.54  &&   2.74 y & 100         &  \\
\hline
$^{54}$Mn & $^{54}$Cr&    5.99  &&  312.1 d & 100         &  100\% (834.9 keV)\\
\hline
$^{51}$Cr & $^{51}$V  &    5.46  &&   27.7 d & 100         &  9.9\% (320.1 keV)\\
\hline
$^{49}$V   & $^{49}$Ti&    4.97  & &  330 d & 100         &  \\
\hline
$^{44}$Ti & $^{44}$Sc &    4.49  &&   60.0 y & 100         &  100\% (78.3 + 67.9 keV)\\
$^{44}$Sc & $^{44}$Ca &    4.04  &&   4.0 h & 100        &  99.9\% (1157.0 keV)\\
\bottomrule
\end{tabular}
\end{center}
\end{table}

\begin{table}
\caption{Measured decay rates of different isotopes ($\frac{dN}{dt}$) along with their uncertainties ($\sigma(\frac{dN}{dt})$) per detector. Both are given in units of decays$\cdot$kg$^{-1}$day$^{-1}$. In case of the $^{3}$H the error on the measured rate includes statistical and systematic uncertainties. Systematic uncertainty is based on the comparison of the decay rate with the flat/inclined spectrum of Compton electrons. Measured rates of $^{65}$Zn decay include 25\% correction as explained in the text. All the rates have been weighted over the fiducial mass considered in the low threshold WIMP search analysis~\cite{lowmass}.
\label{tab:rates}}
\makebox[\linewidth]{
\begin{tabular}{c|cc|cc|cc|cc|cc}
\toprule
 \multirow{ 2}{*}{\textbf{Detector}} &  \multicolumn{2}{c|}{\textbf{$^{3}$H}}&  \multicolumn{2}{c|}{\textbf{$^{65}$Zn}} &  \multicolumn{2}{c|}{\textbf{$^{55}$Fe}} &  \multicolumn{2}{c|}{\textbf{$^{68}$Ge}}&  \multicolumn{2}{c}{\textbf{$^{49}$V}}  \\
    &             $\frac{dN}{dt}$ &  $\sigma(\frac{dN}{dt})$&         $\frac{dN}{dt}$ &  $\sigma(\frac{dN}{dt})$ &           $\frac{dN}{dt}$ &  $\sigma(\frac{dN}{dt})$ &        $\frac{dN}{dt}$&  $\sigma(\frac{dN}{dt})$&        $\sigma(\frac{dN}{dt})$&  $\sigma(\frac{dN}{dt})$ \\
\hline
FID823	&	0.36	&	0.45	&	0.69	&	0.22	&	0.16	&	0.11	&	1.52	&	0.25	&	0.14	&	0.10\\
FID824	&	0.67	&	0.34	&	1.15	&	0.18	&	0.10	&	0.06	&	1.93	&	0.23	&	0.12	&	0.07\\
FID825	&	0.30	&	0.29	&	1.14	&	0.18	&	0.10	&	0.06	&	2.21	&	0.22	&	0.03	&	0.05\\
FID826	&	0.92	&	0.46	&	1.75	&	0.26	&	0.17	&	0.10	&	2.58	&	0.35	&	0.13	&	0.11\\
FID827	&	1.68	&	0.51	&	3.84	&	0.27	&	0.31	&	0.09	&	8.77	&	0.42	&	0.18	&	0.08\\
FID828	&	0.93	&	0.55	&	3.18	&	0.28	&	0.26	&	0.09	&	11.56	&	0.51	&	0.17	&	0.08\\
FID837	&	0.75	&	0.47	&	3.29	&	0.27	&	0.19	&	0.08	&	7.37	&	0.39	&	0.12	&	0.07\\
FID838	&	1.10	&	0.35	&	3.57	&	0.27	&	0.27	&	0.09	&	11.50	&	0.46	&	0.15	&	0.07\\
FID839	&	2.35	&	0.45	&	2.45	&	0.23	&	0.16	&	0.08	&	6.84	&	0.36	&	0.04	&	0.07\\
FID841	&	1.18	&	0.42	&	3.21	&	0.26	&	0.12	&	0.07	&	4.33	&	0.29	&	0.12	&	0.07\\
FID842	&	2.09	&	0.50	&	3.03	&	0.27	&	0.38	&	0.11	&	4.25	&	0.31	&	0.09	&	0.08\\
FID844	&	1.07	&	0.41	&	1.21	&	0.21	&	0.12	&	0.08	&	3.93	&	0.29	&	0.12	&	0.08\\
FID845	&	2.52	&	0.86	&	1.44	&	0.23	&	0.21	&	0.11	&	2.54	&	0.27	&	0.28	&	0.12\\
\bottomrule
\end{tabular}}
\end{table}

\begin{table}\caption{Exposure and cool-off times in days for all detectors: t$_{dec2}$ is calculated as the time between the day when the detector has last been moved underground and the middle of the run. Listed as $\Delta$t is the uncertainty considered in our history model due to the absence of some records.
\label{tab:history}}
\begin{center}
\begin{tabular}{c|cc|cc|cc|cc}
\toprule
Detector & t$_{exp1}$ & $\Delta$t$_{exp1}$ & t$_{dec1}$ & $\Delta$t$_{dec1}$  & t$_{exp2}$ & $\Delta$t$_{exp2}$ & t$_{dec2}$ &$\Delta$t$_{dec2}$ \\
 & \multicolumn{2}{c|}{\footnotesize (days)} &\multicolumn{2}{c|}{\footnotesize (days)} &\multicolumn{2}{c|}{\footnotesize (days)} &\multicolumn{2}{c}{\footnotesize (days)} \\
\hline
 FID823 &   47 &  $\pm$7 & 398 & $\mp$14 &   29 &$\pm$14 & 1002 & $\mp$7\\
 FID824 &   47 & $\pm$7&  404 & $\mp$14&   23 & $\pm$14& 1002& $\mp$7\\
 FID825 &   32 & $\pm$7&  376 & $\mp$14&   20 & $\pm$14& 1002& $\mp$7\\
 FID826 &   58 & $\pm$7&  316 & $\mp$14&   21 & $\pm$14&  984& $\mp$7\\
 FID827 &   82 & $\pm$7 &  110 & $\mp$14&   127 & $\pm$42&  798& $\mp$35\\
 FID828 &   37 & $\pm$7&  123 & $\mp$14&   119 & $\pm$42&  798& $\mp$35\\
  FID837 &   50 & $\pm$7&  394 & $\mp$14&   91 & $\pm$14&  612& $\mp$7\\
 FID838 &   59 & $\pm$7&  342 & $\mp$14&   88 & $\pm$14&  612& $\mp$7\\
 FID839 &   43 & $\pm$7&  347 & $\mp$14&   84 & $\pm$14&  612& $\mp$7\\
 FID841 &   31 & $\pm$7&  819 & $\mp$14&   45 & $\pm$14&  612& $\mp$7\\
 FID842 &   43 & $\pm$7&  821 & $\mp$14&   35 & $\pm$14&  612& $\mp$7\\
 FID844 &   35 & $\pm$7&  833 & $\mp$14&   31 & $\pm$14&  612& $\mp$7\\
 FID845 &   30 & $\pm$7&  831 & $\mp$14&   26 & $\pm$14&  612& $\mp$7\\
  \bottomrule
\end{tabular}
\end{center}
\end{table}

\begin{table}\caption{Saturation fraction $f_{s}$ along with their uncertainties, $\sigma$(f$_{s}$), per detector. \label{tab:saturationfraction}}
\makebox[\linewidth]{
\begin{tabular}{c|C{1cm}C{1.1cm}|C{1cm}C{1.1cm}|C{1cm}C{1.1cm}|C{1cm}C{1.1cm}|C{1cm}C{1.1cm}}
\toprule
%\begin{tabular}{|c|cc|cc|cc|cc|cc|}\hline
 \multirow{ 2}{*}{\textbf{Detector}} &  \multicolumn{2}{c|}{\textbf{$^{3}$H}}&  \multicolumn{2}{c|}{\textbf{$^{65}$Zn}} &  \multicolumn{2}{c|}{\textbf{$^{55}$Fe}} &  \multicolumn{2}{c|}{\textbf{$^{68}$Ge}}&  \multicolumn{2}{c}{\textbf{$^{49}$V}}  \\
    &            $f_{s}$    & $\sigma$($f_{s}$)&     $f_{s}$        &  $\sigma$($f_{s}$) &        $f_{s}$     & $\sigma$($f_{s}$) &    $f_{s}$       &  $\sigma$($f_{s}$) &         $f_{s}$    &  $\sigma$($f_{s}$) \\
\hline
FID823	&	0.0097	&	0.0028	&	0.0068	&	0.0026	&	0.0220	&	0.0068	&	0.0085	&	0.0032	&	0.0119	&	0.0043\\
FID824	&	0.0089	&	0.0028	&	0.0059	&	0.0027	&	0.0200	&	0.0068	&	0.0074	&	0.0032	&	0.0104	&	0.0043\\
FID825	&	0.0067	&	0.0028	&	0.0049	&	0.0028	&	0.0154	&	0.0068	&	0.0061	&	0.0033	&	0.0085	&	0.0044\\
FID826	&	0.0101	&	0.0028	&	0.0071	&	0.0029	&	0.0232	&	0.0070	&	0.0090	&	0.0035	&	0.0126	&	0.0046\\
FID827	&	0.0279	&	0.0068	&	0.0423	&	0.0131	&	0.0755	&	0.0194	&	0.0494	&	0.0150	&	0.0616	&	0.0180\\
FID828	&	0.0209	&	0.0068	&	0.0349	&	0.0131	&	0.0579	&	0.0194	&	0.0404	&	0.0150	&	0.0497	&	0.0180\\
FID837	&	0.0192	&	0.0029	&	0.0459	&	0.0072	&	0.0561	&	0.0086	&	0.0507	&	0.0080	&	0.0580	&	0.0091\\
FID838	&	0.0200	&	0.0029	&	0.0469	&	0.0074	&	0.0584	&	0.0087	&	0.0519	&	0.0082	&	0.0597	&	0.0093\\
FID839	&	0.0174	&	0.0029	&	0.0433	&	0.0074	&	0.0515	&	0.0087	&	0.0477	&	0.0082	&	0.0544	&	0.0093\\
FID841	&	0.0102	&	0.0029	&	0.0225	&	0.0070	&	0.0279	&	0.0082	&	0.0246	&	0.0077	&	0.0279	&	0.0086\\
FID842	&	0.0102	&	0.0029	&	0.0185	&	0.0071	&	0.0265	&	0.0082	&	0.0205	&	0.0078	&	0.0237	&	0.0087\\
FID844	&	0.0087	&	0.0029	&	0.0164	&	0.0072	&	0.0227	&	0.0082	&	0.0180	&	0.0078	&	0.0208	&	0.0087\\
FID845	&	0.0074	&	0.0029	&	0.0139	&	0.0072	&	0.0193	&	0.0083	&	0.0153	&	0.0078	&	0.0176	&	0.0088\\

\bottomrule
\end{tabular}}

\end{table}

\begin{table}
\caption{Table of production rates (fit values) considering  a selection of 13 detectors. No systematics uncertainties are considered. For $^{68}$Ge, a lower limit at saturation (sat) is quoted as well. 
\label{tab:pr}}
\begin{center}
\begin{tabular}{ccc}\hline
Isotope &  \textbf{nuclei/kg/day} & \textbf{$\chi^{2}/d.o.f.$}  \\
\toprule
\textbf{$^{3}$H} & 82 $\pm$  12 &1.76\\
\textbf{$^{49}$V}&2.8 $\pm$  0.6&0.92 \\
\textbf{$^{65}$Zn} &106 $\pm$  10&2.43 \\
\textbf{$^{55}$Fe} & 4.6 $\pm$  0.7 &0.72 \\
\hline
 \multirow{ 2}{*}{\textbf{$^{68}$Ge}} & 202 $\pm$  16 &0.93\\
& 84 $\pm$  3 (sat) &3.17\\
 \bottomrule
\end{tabular}
\end{center}
\end{table}

\begin{table}
\caption{Rates of production (expressed in kg$^{-1}\cdot$ day$^{-1}$) of isotopes induced in natural germanium at sea level as measured in EDELWEISS III germanium detectors, compared with previous estimates and measurements in Ref.~\cite{ceb}, \cite{mei}, ~\cite{Zhang:2016rlz}, \cite{KK} and \cite{avignone}. Errors on the production rate include statistical and for $^{3}$H and $^{65}$Zn systematic uncertainties, too. Systematic uncertainty is based on the minimization of the reduced $\chi^{2}$. Estimate in this work refers to ACTIVIA calculation, considering semi-empirical~\cite{silb,silb1,silb2,silb3,silb4} (a) and MENDL-2P database (b)~\cite{mendl}  cross sections. For $^{49}$V, both calculations give the same result. An upper limit for $^{3}$H from IGEX data (E) is shown together with calculations~\cite{mei} for all isotopes. (I) and (II) refer to GEANT4 and ACTIVIA calculations from~\cite{Zhang:2016rlz}. The lower limit for $^{68}$Ge at saturation value is listed. It is derived from the fit value of 84 $\pm$ 6  at a 90\%~C.L.\. The last two columns refer to estimates from model~\cite{Hess} and experimental data (Exp.) from Ref.~\cite{avignone}. 
\label{tab:results}}
 \makebox[\linewidth]{
\begin{tabular}{C{.9cm} C{2.cm}C{1cm}C{1cm}C{1cm}C{1.8cm}C{1.3cm}C{1.2cm}C{1.1cm}C{1.1cm}}\hline
\toprule
%\begin{tabular*}{\textwidth}{c @{\extracolsep{\fill}} ccccccccc}
  &   \multicolumn{2}{c}{This work}  & \multicolumn{2}{c}{{\small Ref.~\cite{ceb}}}   & \multirow{ 2}{*}{{\small Ref.~\cite{mei}}} & \multirow{ 2}{*}{{\small Ref.~\cite{Zhang:2016rlz}}}& \multirow{ 2}{*}{{\small Ref.~\cite{KK}}} & \multicolumn{2}{c}{{\small Ref.~\cite{avignone}}} \\
% & & & (Ziegler et al.) & (Gordon et al.) &&& (MC)& {exp}\\
  & Exp. & Calc.& {\tiny (Ziegler)} & {\tiny (Gordon)} &&&& {\footnotesize From~\cite{Hess}}& {\footnotesize Exp.}\\
\hline
\multirow{ 2}{*}{\textbf{$^{3}$H}} &  \multirow{ 2}{*}{82$\pm$21} &  46{\tiny (a)}&&&27.7&48.3 {\tiny(I)}&&\multirow{ 2}{*}{210}\\
&&43.5{\tiny(b)}&&&<21{\tiny (E)}&52.4 {\tiny(II)}&&\\
\textbf{$^{49}$V} & 2.8$\pm$0.6 & 1.9 (a,b) &  & & & && &\\
\multirow{ 2}{*}{\textbf{$^{65}$Zn}} &  \multirow{ 2}{*}{106$\pm$13} & 38.7{\tiny(a)} & \multirow{ 2}{*}{77} & \multirow{ 2}{*}{63} &\multirow{ 2}{*}{37.1} && \multirow{ 2}{*}{79} & \multirow{ 2}{*}{34.4} &\multirow{ 2}{*}{38$\pm$6}\\
&&65.8{\tiny(b)}&&&&&\\
\multirow{ 2}{*}{\textbf{$^{55}$Fe}} & \multirow{ 2}{*}{4.6$\pm$0.7}  & 3.5{\tiny(a)} & \multirow{ 2}{*}{8.0} & \multirow{ 2}{*}{6.0} & \multirow{ 2}{*}{8.6} && \multirow{ 2}{*}{8.4}  & & \\
&&4.0{\tiny(b)}&&&&&\\
\hline
\multirow{ 2}{*}{\textbf{$^{68}$Ge}} & \multirow{ 2}{*}{>74}   & 23.1{\tiny(a)}& \multirow{ 2}{*}{89} &\multirow{ 2}{*}{60} & \multirow{ 2}{*}{41.3}&&\multirow{ 2}{*}{58.4} & \multirow{ 2}{*}{29.6}&\multirow{ 2}{*}{30$\pm$7} \\
&  & 45.0{\tiny(b)}& &&  &&&\\
\bottomrule
\end{tabular}}
\end{table}

\end{document}